\documentclass[twocolumn,showpacs,preprintnumbers,amsmath,amssymb,superscriptaddress, prl,floatfix]{revtex4-1}
\usepackage{graphicx}% Include figure files
\usepackage{dcolumn}% Align table columns on decimal point
\usepackage{tabularx}
\usepackage{bm}% bold math
\usepackage{epstopdf}
\usepackage{amsmath}
\usepackage{color}
\usepackage{here}
\usepackage{float}

\usepackage{titlesec}
\titlespacing*{\section}{0pt}{1.1\baselineskip}{\baselineskip}

\begin{document}

\preprint{Preprint}

\title{2D Fourier Transform Spectroscopy of exciton-polaritons and their interactions}% Force line breaks with \\

\author{N. Takemura}
\email[E-mail: ]{naotomo.takemura@epfl.ch}
\affiliation{Laboratory of Quantum Optoelectronics, \'Ecole Polytechnique F\'ed\'erale de Lausanne, CH-1015, Lausanne, Switzerland}
\author{S. Trebaol}
\affiliation{UMR FOTON, CNRS, Universit\'e de Rennes 1, Enssat, Insa Rennes, 6 rue de Kerampont 22305 Lannion, France}
\author{M. D. Anderson}
\affiliation{Laboratory of Quantum Optoelectronics, \'Ecole Polytechnique F\'ed\'erale de Lausanne, CH-1015, Lausanne, Switzerland}
\author{V. Kohnle}
\affiliation{Laboratory of Quantum Optoelectronics, \'Ecole Polytechnique F\'ed\'erale de Lausanne, CH-1015, Lausanne, Switzerland}
\author{Y. L\'eger}
\affiliation{UMR FOTON, CNRS, Universit\'e de Rennes 1, Enssat, Insa Rennes, 6 rue de Kerampont 22305 Lannion, France}
\author{ D. Y. Oberli}
\affiliation{Laboratory of Quantum Optoelectronics, \'Ecole Polytechnique F\'ed\'erale de Lausanne, CH-1015, Lausanne, Switzerland}
\author{M. T. Portella-Oberli}
\affiliation{Laboratory of Quantum Optoelectronics, \'Ecole Polytechnique F\'ed\'erale de Lausanne, CH-1015, Lausanne, Switzerland}
\author{B. Deveaud}
\affiliation{Laboratory of Quantum Optoelectronics, \'Ecole Polytechnique F\'ed\'erale de Lausanne, CH-1015, Lausanne, Switzerland}

\date{\today}% It is always \today, today,

\begin{abstract}
We investigate polariton-polariton interactions in a semiconductor microcavity through two-dimensional Fourier transform (2DFT) spectroscopy. We observe, in addition to the lower-lower and the upper-upper polariton self-interaction, a lower-upper cross-interaction. This appears as separated peaks in the on-diagonal and off-diagonal part of 2DFT spectra. Moreover, we elucidate the role of the polariton dispersion through a fine structure in the 2DFT spectrum. Simulations, based on lower-upper polariton basis Gross-Pitaevskii equations including both self and cross-interactions, result in a 2DFT spectra in qualitative agreement with experiments. 
\end{abstract}

%68.35.-p 	Solid surfaces and solid?solid interfaces: structure and en3333ergetics
%68.37.Ma 	Scanning transmission electron microscopy (STEM) 
%79.20.Uv 	Electron energy loss spectroscopy

\maketitle

\section{I. INTRODUCTION}
The strong coupling between quantum well excitons and photons confined in a microcavity gives rise to two new eigenstates: lower and upper-polaritons. Furthermore, polariton-polariton interactions (anharmonicities), mediated by the nonlinear interaction of excitons, provide a wide range of rich physics. In fact, the lower and upper polariton states are no more exact eigenstates, because they are defined in a non-interacting (harmonic) exciton system \cite{Hopfield1958}. The exciton-exciton interaction introduces not only lower and upper-polariton self-interactions but also lower-upper cross-interactions. In quantum chemistry, this type of problem is known as ``normal mode versus local mode" problem \cite{Hamm2011,GuidoDellaValle1988,Gorbunov2005}. In our context, the local and normal modes are respectively the exciton-photon and lower-upper polariton states. Although a wide range of research has been made on lower polaritons, such as superfluity \cite{Amo2009} and its Bogoliubov excitation spectrum \cite{Kohnle2011,Utsunomiya2008}, the self and cross-interaction between lower and upper-polaritons has not yet been fully investigated. In fact, it is difficult to distinguish lower-upper polariton cross-interaction from self-interactions through conventional one-dimensional (1D) spectroscopy \cite{Takemura2014}. In order to enlighten the polariton interactions, it is useful to employ a two-dimensional Fourier transform (2DFT) spectroscopy technique. 

2DFT spectroscopy is a powerful tool to investigate coherent couplings and vibrational anharmonicities of molecular vibrational states \cite{Golonzka2001,Hamm2011}. One advantage of a 2D spectrum is that we can associate each peak of the spectrum with different Liouville-space pathway through double-sided Feynman diagrams \cite{Hamm2011,Khalil2001}. With this idea, we can identify dominant nonlinear optical pathways in four-wave mixing (FWM) signals, which we cannot access with conventional 1D spectra. This method has been extended to investigations of electron-hole many-body properties in semiconductor systems \cite{Li2006,Stone2009,Kasprzak2011,Nardin2014}. Those researches revealed the importance of exciton-exciton interactions, excitation-induced dephasing (EID), and bound biexcitons in quantum wells. Recently, the 2DFT spectroscopy technique has been applied to semiconductor microcavity polaritons \cite{1367-2630-15-2-025005}. 

In this paper, we report on 2DFT spectra when both lower and upper polariton states are simultaneously excited. We use 2DFT spectroscopy to differentiate the two types of nonlinearities: self and cross-polariton interactions. We perform two-pulse FWM experiments in both positive and negative time delay configurations. Polaritons inherit, from their photonic component, a light effective mass that leads to a strong parabolicity in energy-momentum dispersion, which is generally neglected in bare quantum well excitons. We reveal the role played by the energy-momentum dispersion on the nonlinear polariton dynamics, which is usually not involved in 2DFT spectroscopy of heavy particles. This paper is organized as follows: Section II, describes the sample and the four-wave mixing experiment, Section III reports on the experimental results and a simple third-order perturbative analysis, and Section IV is dedicated to a detailed theoretical model and numerical simulation using lower- and upper-polariton basis Gross-Pitaevskii equations.
 
\section{II. EXPERIMENTAL METHOD}
\begin{figure}[h]
\includegraphics[width=0.4\textwidth]{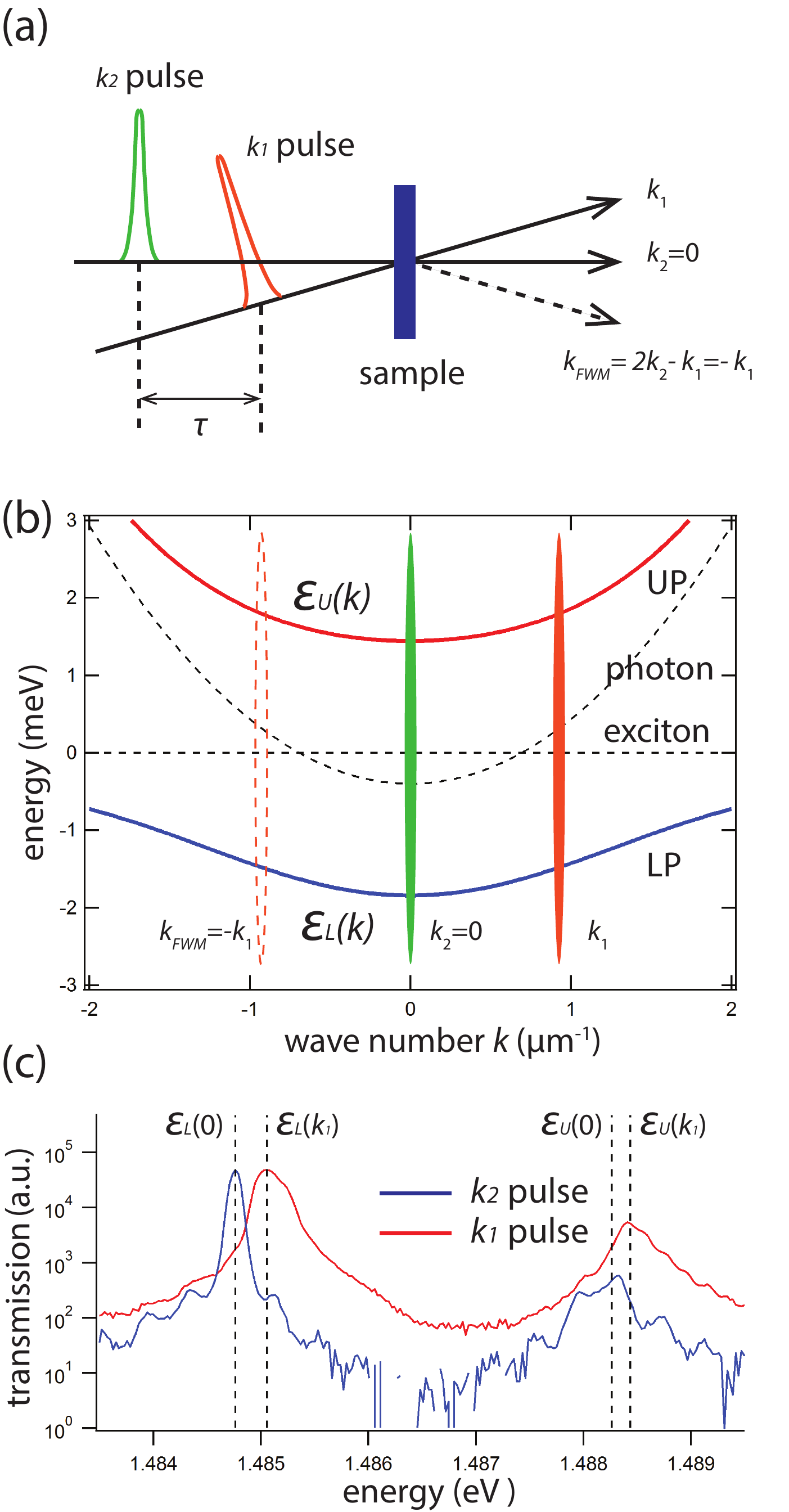}
\caption{(color online) Schematic of FWM configuration and pulse sequence (a). Lower- (LP) and upper-polariton (UP) energy-momentum dispersion at a slight negative cavity detuning (b). The dashed black lines represent exciton and photon and energy-momentum dispersion.  Transmission spectrum of the $k_2$ and $k_1$ beam (c). $\epsilon_L(k)$ and $\epsilon_U(k)$ are respectively lower and upper-polariton enegy at wave-vector $k$}
\label{fig:schematic}
\end{figure}

The sample is a high quality III-V GaAs-based micro-cavity \cite{Stanley1994}. A single 8 nm In$_{0.04}$Ga$_{0.96}$As quantum well is sandwiched between a pair of GaAs/AlAs distributed Bragg-reflectors. The Rabi splitting energy at zero cavity-exciton detuning ($\delta$=0) is $\Omega=$3.26 meV. The experiments are performed at the cryogenic temperature of 4 K with several positions on the sample corresponding to different exciton-cavity detunings. We use a Ti:sapphire laser with a broad spectrum femtosecond pulse and 80 MHz repetition rate. The center energy of the pulse spectrum is set between the lower and upper polariton energies. We employ four-wave-mixing spectroscopy in two-pulses configuration. The $k_2$ and $k_1$ pulses arrive on the sample in directions $\vec{k}_1 = 0.96$ $\mu$m$^{-1}$ and $\vec{k}_2$=0 $\mu$m$^{-1}$ respectively (See Fig. \ref{fig:schematic} (a)). The experiments are performed in the low-density regime with $k_1$ and $k_2$ pulse intensities of $6.7\times 10^{12}$ photons pulse$^{-1}$ cm$^{-2}$. The pulses are co-circularly polarized in order to avoid the biexciton effect \cite{Takemura2014a,Takemura2014}. We detect the FWM signal in the direction $\vec{k}_{FWM}=2\vec{k}_2-\vec{k}_1 = -\vec{k}_1$. The pulses $k_1$ and $k_2$ arrive on the sample at times $t_{k_1}$ and $t_{k_2}$ respectively. A time delay $\tau=t_{k_1}-t_{k_2}$ between two pulses is called positive (negative) when the $k_2$ ($k_1$) pulse arrives before the $k_1$ ($k_2$) pulse. In Fig. \ref{fig:schematic} (b), we show schematically the polariton dispersion with the $k_1$ and $k_2$ pulses and also the $k_{\rm FWM}$ signal. Fig. \ref{fig:schematic}(c) shows that the transmissions of $k_2$ and $k_1$ pulse have different energy peaks due to the effect of the polariton energy dispersion.  

The experimental set up is explained in detail in our previous work \cite{Kohnle2012}. With a heterodyne detection technique \cite{Hall1992}, we record the electric field of the FWM signal $S(\tau,t)$, where $t$ represents the real evolution time of the FWM signal after the incidents of the two pulses. Notice that the FWM signal $S(\tau,t)$ is function of two-independent time periods, the time delay $\tau$ and the real time evolution $t$ of the signal. We obtain a delay dependent 1D FWM amplitude signal spectrum $|S(\tau,\epsilon_t)|$ by performing a Fourier transformation (FT) with respect to $t$, which is performed by spectral interferometry \cite{Kohnle2011}. The spectrometer acts as a FT, converting the real time evolution $t$ into the third-order emission energy $\epsilon_t$. The 2D spectrum $S(\epsilon_\tau,\epsilon_t)$ is then obtained through the FT with respect to both the $\tau$ and $t$ axes. Here $\epsilon_\tau$ and $\epsilon_t$ represent the absorption and emission energies respectively. Before the conversion to the 2DFT spectrum, we apply a numerical phase correction \cite{Albert2013} using the upper-polariton energy as a phase reference (a detailed explanation is given in Appendix A).

\section{III. EXPERIMENTAL RESULTS AND PERTURBATIVE ANALYSIS}
In this section, we present experimental results from FWM experiments performed with the cavity detuning at $\delta=-0.38$ meV and analyse them with a conventional third-order perturbation theory of nonlinear optics. The detail of the third-order perturbation theory is presented in Appendix B.

\subsection{A. 1D FWM spectrum}
In Fig. \ref{fig:exp_PRB} (a), we display the amplitude of the 1D FWM signal $|S(\tau,\epsilon_t)|$, which is the spectrum of the emitted signal as function of the delay time, $\tau$, between the two pulses. The 1D FWM signal spectrum presents two main resonances: the lower frequency emission, originating from lower polariton (LP), and the higher frequency emission, from upper polariton (UP). Each one displaying a fine structure. Moreover, the FWM emission shows a temporal oscillation behaviour with a period of 1.2 ps. This period corresponds to the Rabi splitting energy. We will show that this oscillation can be understood as a quantum beat. All these features are addressed in the next sections.

\subsection{B. 2DFT spectrum}
The absolute value of the 2DFT spectrum $|S(\epsilon_\tau,\epsilon_t)|$ is shown in Fig. \ref{fig:exp_PRB} (b) for negative $\tau$ and in Fig. \ref{fig:exp_PRB} (c) for positive $\tau$. Fig. \ref{fig:exp_PRB} (b) and (c) are respectively referred to as the one-quantum and two-quantum regimes \cite{Hamm2011}. A Fourier transformation with respect to the time delay $|\tau|$ converts the delay map $S(|\tau|,\epsilon_t)$ into the 2DFT spectrum $S(\epsilon_\tau,\epsilon_t)$. One axis displays the absorption energy $\epsilon_\tau$ and the other the emission energy $\epsilon_t$. We will analyse the 2DFT spectrum in terms of third-order perturbation theory with the help of double-sided Feynman diagrams. 

\subsection{C. 2DFT: One-quantum regime}
\begin{figure}[h]
\includegraphics[width=0.35\textwidth]{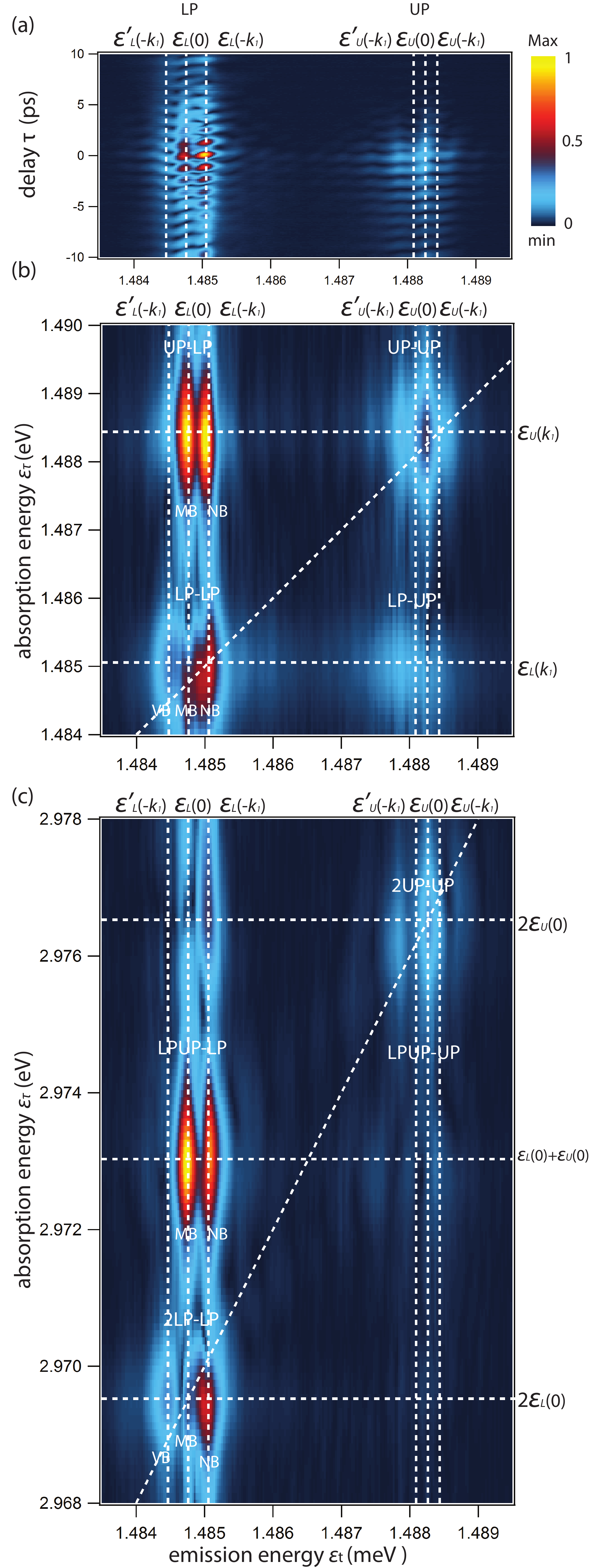}
\caption{(color online) Experimental amplitude of FWM spectrum as a function of emission energy and $k_2$-$k_1$ pulse delay $\tau$: $|S(\tau,\epsilon_t)|$ (a). Amplitude of 2DFT spectrum $|S(\epsilon_\tau,\epsilon_t)|$ for 1-quantum (b) and 2-quantum region (c). Diagonal dashed lines represent $\epsilon_\tau=\epsilon_t$ for (b) and $\epsilon_\tau=2\epsilon_t$ for (c). Horizontal and vertical dashed lines respectively represent different absorption and emission energies. $\epsilon'_{L(U)}(k)$ is a virtual branch (VB), which is explained in Section IV. c. Colour scales are normalized by the maximum and minimum of the amplitude. The same normalization and colour bar are used in all figures of the article.}
\label{fig:exp_PRB}
\end{figure}

Firstly, we focus on the FWM signal emitted at the negative time delay ($\tau<0$). In Fig. \ref{fig:exp_PRB} (b), we observe two diagonal groups, LP-LP and UP-UP, and two off-diagonal groups, LP-UP and UP-LP. Inside each peak group, fine structures are found. We classify and name them the virtual (VB), middle (MB), and normal (NB) branches going from lower to higher emission energies. In order to analyse the origin of LP-LP and UP-LP groups, we introduce double-sided Feynman diagrams  (Fig. \ref{fig:negative_doublesided_PRB}) \cite{Mukamel1995,Hamm2011,Yang2007}. The double-sided Feynman diagrams represent Liouville-space pathways of the FWM signals. Each diagram corresponds to the third-order perturbative evolution of the system's density matrix. The FWM signal $S(|\tau|,t)$ can be calculated, within third-order perturbation, by summing all pathways. In the diagrams, the vertical line is the time evolution of the system, with the time ordering of the arrival of $k_1$, $k_2$ pulses and the FWM signal emission time. In all pathways, the arrival of the first $k_1$ pulse creates a coherence between the ground state and a single LP (0-LP) or UP (0-UP) state. Both the second and third fields come from the pulse $k_2$ after a delay $|\tau|$ simultaneously, which fixes $T=0$ in the diagrams. Performing a standard third-order perturbative calculation of nonlinear optics \cite{Hamm2011,Khalil2001} (See the Appendix B), the FWM signal corresponding to the diagram (A) is given by,
\begin{eqnarray}\nonumber
S^{(A)}(\tau,t)&\propto&|\Omega_{L}|^4e^{-(i/\hbar)(\epsilon_{L}(\vec{k}_{FWM})-i\gamma_L)t}e^{(i/\hbar)(\epsilon_{L}(\vec{k}_1)+i\gamma_L)|\tau|}.
\end{eqnarray}
where $\epsilon_L(k)(\epsilon_U(k))$ is the energy of the lower (upper)-polariton, $\Omega_L(\Omega_U)$ represents the coupling constant between the lower (upper)-polariton and the photon outside the cavity, and $\gamma_L(\gamma_U)$ is the dephasing rate of lower (upper)-polariton. In this pathway, during $|\tau|$ and $t$, the system evolves keeping a coherence between its ground state and a single lower-polariton state (0-LP and LP-0). In the ``one-quantum regime", the first order evolution, during time $|\tau|$, is always a coherence between the ground state and single lower or upper-polariton state (0-LP or 0-UP). The Fourier transformation of $S^{(A)}(|\tau|,t)$ reads,
\begin{eqnarray}\nonumber
S^{(A)}(\epsilon_\tau,\epsilon_t)
&\propto&\frac{|\Omega_{L}|^4}{[i(\epsilon_t-\epsilon_L(\vec{k}_{FWM}))+\gamma_L][i(\epsilon_\tau+\epsilon_L(\vec{k}_1))+\gamma_L]}.
\end{eqnarray}
Similarly, the contributions from diagrams (B) and (C) lead to
\begin{eqnarray}\nonumber
S^{(B)}(\epsilon_\tau,\epsilon_t)
&\propto&\frac{|\Omega_{L}|^4}{[i(\epsilon_t-\epsilon_L(\vec{k}_{FWM}))+\gamma_L][i(\epsilon_\tau+\epsilon_L(\vec{k}_1))+\gamma_L]},
\end{eqnarray}
and
\begin{eqnarray}\nonumber
&S^{(C)}&(\epsilon_\tau,\epsilon_t)\\\nonumber
&\propto&-2|\Omega_{L}|^4\cdot\frac{1}{[i(\epsilon_t-\epsilon_{2L}(\vec{k}_{FWM})+\epsilon_{L}(\vec{k}_{FWM}))+\gamma_L]}\\\nonumber
& &\frac{1}{[i(\epsilon_\tau+\epsilon_L(\vec{k}_1))+\gamma_L]}.
\end{eqnarray}
\begin{figure}[h]
\includegraphics[width=0.35\textwidth]{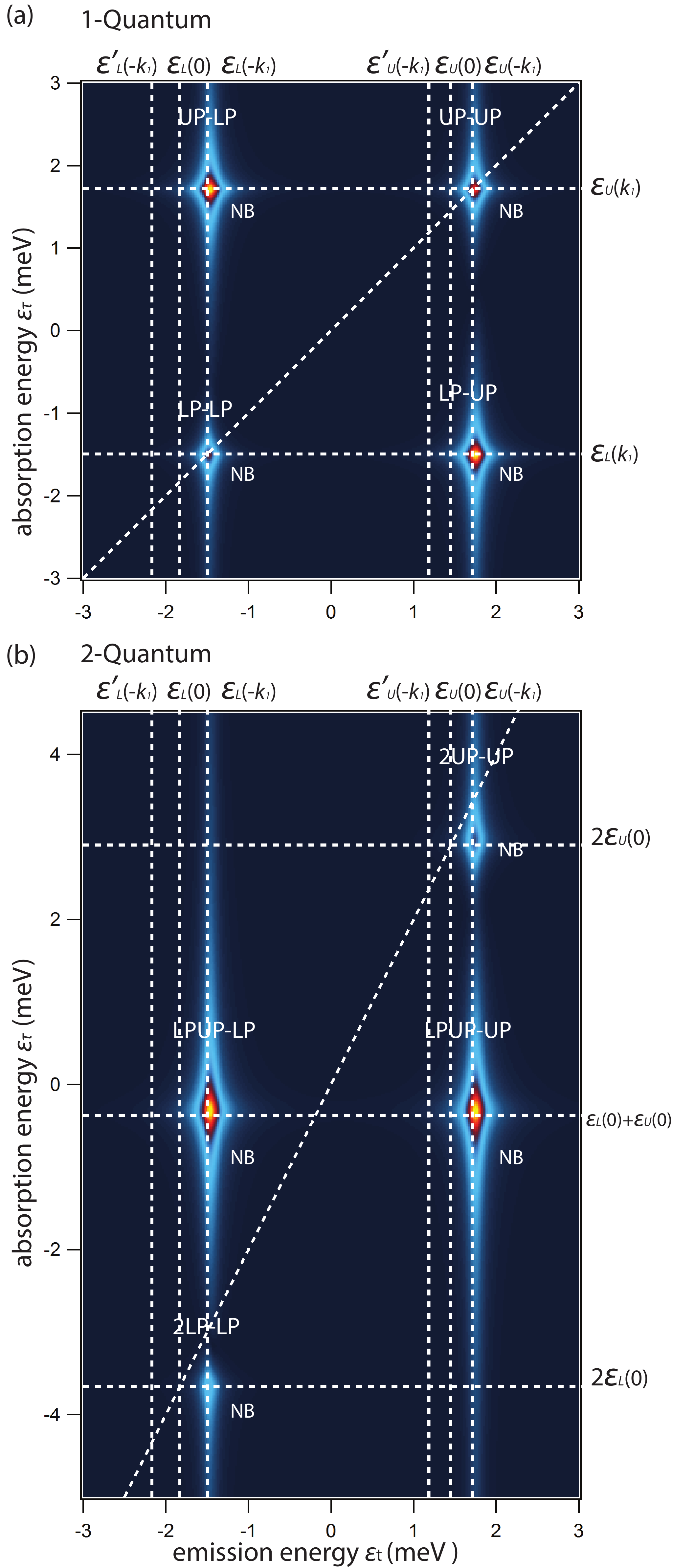}
\caption{(color online) Plot of the amplitude of 2DFT spectrum $|S(\epsilon_\tau,\epsilon_t)|$ based on third-order perturbative calculations for the (a) 1-quantum and (b) 2-quantum regimes.}
\label{fig:sim_analytical}
\end{figure}
$S^{(B)}$ has the exactly same form as $S^{(A)}$. Here, we treat the semiconductor microcavity system as two-oscillators (lower and upper-polaritons) weakly coupled to photons outside the cavity with the couplings $\Omega_L$ and $\Omega_U$. The detailed background of this polariton basis model will be discussed in the theoretical model section. Note that diagrams (A) and (B) in Fig. \ref{fig:negative_doublesided_PRB} include only the ground and ``single-quantum" state (LP), while diagram (C) also includes the ``two-quantum" state (2LP). This two-quantum state (2LP) is modified by the polariton-polariton self-interaction, resulting in the energy of 2LP state $\epsilon_{2L}(\vec{k})$ being slightly blue shifted from twice that of LP state $2\epsilon_{L}(\vec{k})$ (\textit{i.e.,} $\epsilon_{2L}(\vec{k})\neq 2\epsilon_{L}(\vec{k})$). It is worth noting, that if the lower-polariton self-interaction were absent, the relation $\epsilon_{2L}(\vec{k})=2\epsilon_{L}(\vec{k})$ would hold and the sum $S^{(A)}+S^{(B)}+S^{(C)}$ would be zero. This is an intuitive consequence of the fact that no FWM signal appears in a linear system \cite{footnote}. The same description and pathways are applied to the UP-UP resonance, considering, in this case, only the upper polariton coherence and the upper-polariton self-interaction.
\begin{figure}[h]
\includegraphics[width=0.37\textwidth]{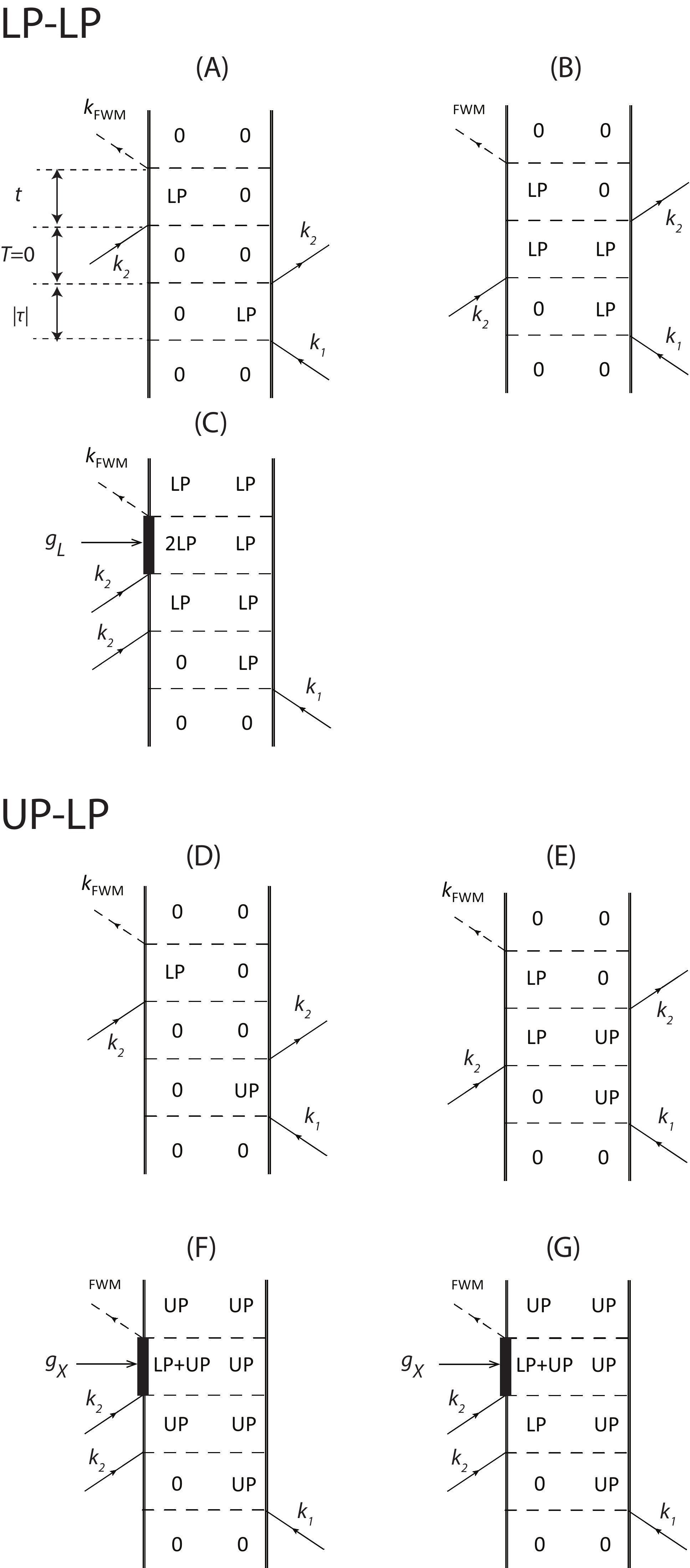}
\caption{Double sided Feynman diagrams that represent perturbative Liouville-space pathways of 1-quantum 2D FT spectrum. Diagrams (A-C) and (D-G) respectively represent the LP-LP and UP-LP peaks (Fig. \ref{fig:exp_PRB} (b)). Polariton-polariton interactions are introduced in pathways (C), (F), and (G), where double $k_2$ pulses excite two-polariton state. $g_L$ and $g_X$ respectively represent self and cross-interaction of polaritons.}
\label{fig:negative_doublesided_PRB}
\end{figure}

Similar to the LP-LP group, we present the Liouville-space pathway processes of UP-LP FWM signals in terms of double-sided Feynman diagrams in Fig. \ref{fig:negative_doublesided_PRB} (D)-(G). The signals associated with these diagrams are given by,
\begin{eqnarray}\nonumber
S^{(D,E)}(\epsilon_\tau,\epsilon_t)
&\propto&\frac{|\Omega_{L}|^2|\Omega_{U}|^2}{[i(\epsilon_t-\epsilon_L(\vec{k}_{FWM}))+\gamma_L][i(\epsilon_\tau+\epsilon_U(\vec{k}_1))+\gamma_U]},
\end{eqnarray}
and
\begin{eqnarray}\nonumber
&S^{(F,G)}&(\epsilon_\tau,\epsilon_t)\\\nonumber
&\propto&-|\Omega_{L}|^2|\Omega_{U}|^2\cdot\frac{1}{[i(\epsilon_t-\epsilon_{LU}(\vec{k}_{FWM})+\epsilon_{U}(\vec{k}_{FWM}))+\gamma_L]}\\\nonumber
& &\frac{1}{[i(\epsilon_\tau+\epsilon_U(\vec{k}_1))+\gamma_U]},
\end{eqnarray}
Along pathways (F) and (G) the FWM emission originates from the coherence between the two-quantum state (LP+UP) and the single-quantum state (UP). Similar to the pathways (A)-(C), the energy of UP-LP state is shifted due to the lower and upper-polariton cross-interaction, $\epsilon_{LU}(\vec{k})\neq\epsilon_{L}(\vec{k})+\epsilon_{U}(\vec{k})$. Again, if the lower and upper-polariton cross-interaction does not exist, $\epsilon_{LU}(\vec{k})=\epsilon_{L}(\vec{k})+\epsilon_{U}(\vec{k})$ holds and the summation of the pathways (D)-(G) cancels. This leads to the disappearance of the off-diagonal peaks. We can draw similar diagrams for LP-UP groups (not shown) and calculate perturbatively the FWM signals. The plot of calculated 2DFT spectrum including all pathways in the one-quantum regime is shown Fig. \ref{fig:sim_analytical} (a).   

As we discussed above, the diagonal peaks arise from the polariton self-interactions while the off-diagonal peaks arise from the cross-interactions. Moreover, the double-sided Feynman diagram analysis elucidates the origin of the amplitude oscillation in the FWM emission along the delay of the two pulses (Fig. \ref{fig:exp_PRB} (a)). This amplitude oscillation can be understood as a quantum beat: an interference of the pathways (A)-(C) and (D)-(G).  During the delay $\tau$, in the pathways (A)-(C) the phase evolves as $e^{(i\epsilon_{L}/\hbar)\tau}$ while in the pathways (D)-(G) it evolves as $e^{(i\epsilon_{U}/\hbar)\tau}$. Thus, the amplitude of the FWM signal of the delay map (Fig. \ref{fig:exp_PRB} (a)) shows a beat frequency which corresponds to the Rabi splitting energy: $\epsilon_{U}-\epsilon_{L}\simeq3.26$ meV.
 
It is important to note that the lower and upper-polariton branches have energy-momentum dispersions: $\epsilon_{\rm L}(\vec{k})\simeq\epsilon_{L,0}+\frac{\hbar^2}{2m_{L}}\vec{k}^2$ and $\epsilon_{\rm U}(\vec{k})\simeq\epsilon_{U,0}+\frac{\hbar^2}{2m_{U}}\vec{k}^2$ (See Fig. \ref{fig:schematic} (b)), where $m_L$ and $m_U$ are the mass of the lower and upper-polaritons respectively. As $\epsilon_{L(U)}(k_{\rm FWM})=\epsilon_{L(U)}(-k_1)=\epsilon_{L(U)}(k_1)$, the LP-LP (UP-UP) peaks are absorbed and emitted at the same energy in the third-order perturbation theory. These peaks presented in Fig. \ref{fig:sim_analytical}(a) correspond to the normal branches (NB) of the experimental 2D spectra in Fig. \ref{fig:exp_PRB} (b). We notice that the third-order perturbative model reproduces only the normal branches, which are resonant to the polariton energy-momentum dispersion, and no fine structure appears inside each group. This is because polariton-polariton interaction is considered only as a level-shift of the eigen state energy in the third-order perturbation theory. In order to give rise to fine energy structures, the superposition between different momentum states induced by polariton-polariton interaction needs to be considered \cite{Savvidis2001}. For doing this, we employ non-perturbative numerical simulations in Section IV. 
\begin{figure}[h]
\includegraphics[width=0.37\textwidth]{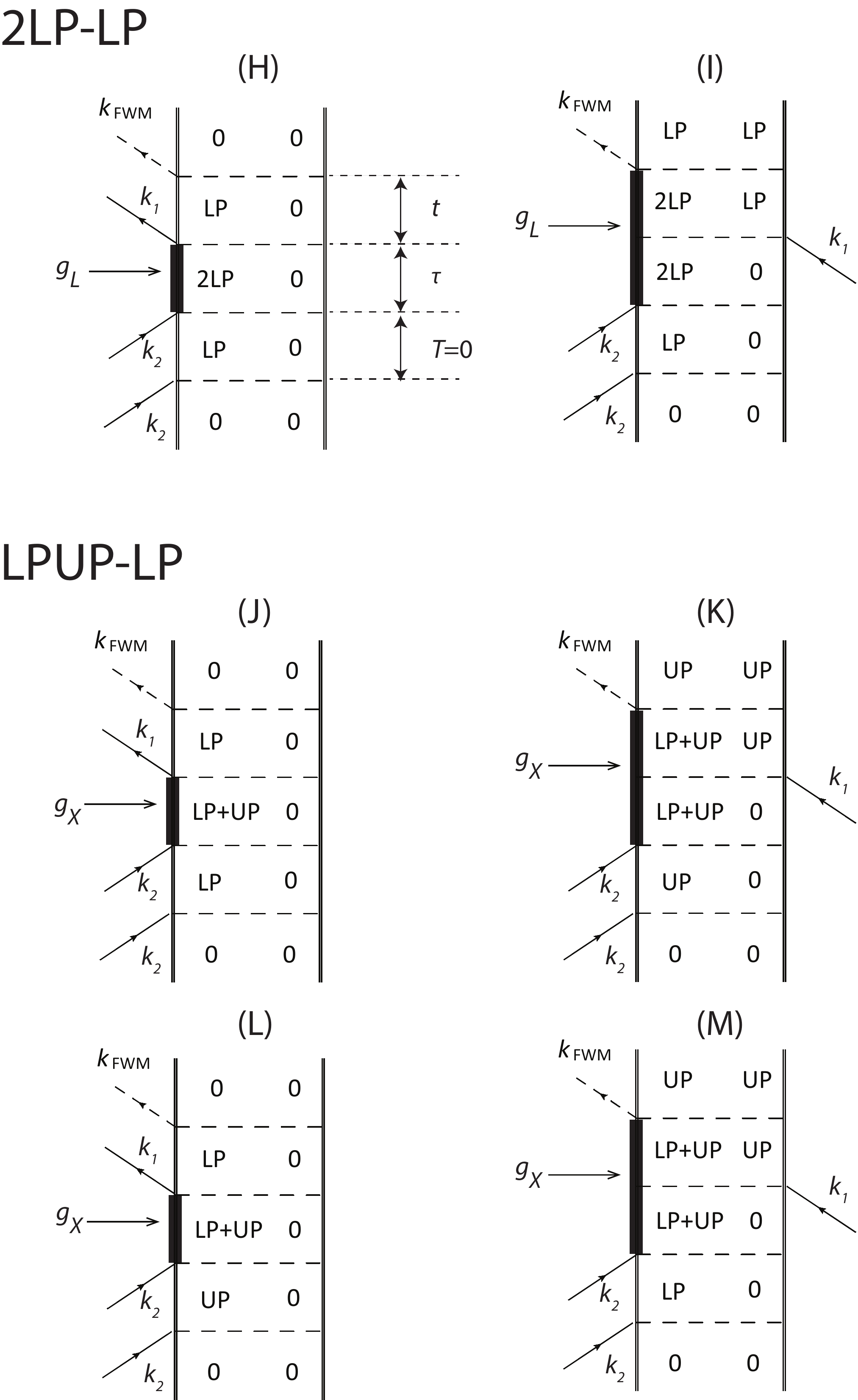}
\caption{Double sided Feynman diagrams that represent perturbative Liouville-space pathways of 2-quantum 2DFT spectrum. Diagrams (H-I) and (J-M) respectively represent the normal branches of 2LP-LP and LPUP-LP group (Fig. \ref{fig:schematic} (b)). Polariton-polariton interactions are present in all pathways. $g_L$ and $g_X$ respectively represent self and cross-interaction of polaritons.}
\label{fig:positive_doublesided_PRB}
\end{figure}

\subsection{D. 2DFT: Two-quantum regime}
We concentrate now on the FWM signal emitted at positive time delays ($\tau>0$), where the $k_2$ pulse arrives first. Since the $k_2$ pulse acts as two degenerate pulses, this pulse creates a coherence between the ground state and the two-quantum state. Thus we call this time delay configuration ``two-quantum regime". Double-sided Feynman diagrams corresponding to 2LP-LP and LPUP-LP groups are presented in Fig. \ref{fig:positive_doublesided_PRB}. In all pathways, the system evolves in two-quantum coherence (0-2LP or 0-LPUP) during time $\tau$, before the arrival of the second pulse $k_1$. For example, the FWM signal corresponding to the diagram (H) can be written as
\begin{eqnarray}\nonumber
S^{(H)}(\tau,t)&\propto&2|\Omega_{L}|^4e^{-(i/\hbar)(\epsilon_{L}(\vec{k}_{FWM})-i\gamma_L)t}e^{-(i/\hbar)(\epsilon_{2L}(\vec{k}_2)-i\gamma_L)\tau}.
\end{eqnarray}
For which the Fourier transformation gives, 
\begin{eqnarray}\nonumber
S^{(H)}(\epsilon_\tau,\epsilon_t)&\propto&2|\Omega_{L}|^4\cdot\frac{1}{[i(\epsilon_t-\epsilon_L(\vec{k}_{FWM}))+\gamma_L]}\\\nonumber
& &\frac{1}{[i(\epsilon_\tau-\epsilon_{2L}(\vec{k}_2))+2\gamma_L]}.
\end{eqnarray}
Similarly,
\begin{eqnarray}\nonumber
&S^{(I)}&(\epsilon_\tau,\epsilon_t)\\\nonumber
&\propto&-2|\Omega_{L}|^4\cdot\frac{1}{[i(\epsilon_t-\epsilon_{2L}(\vec{k}_{FWM})+\epsilon_{L}(\vec{k}_{FWM}))+\gamma_L]}\\\nonumber
& &\frac{1}{[i(\epsilon_\tau-\epsilon_{2L}(\vec{k}_2))+2\gamma_L]}.
\end{eqnarray}
For LPUP-LP groups, the FWM contributions read,
\begin{eqnarray}\nonumber
S^{(J)}(\epsilon_\tau,\epsilon_t)
&\propto&|\Omega_{L}|^2|\Omega_{U}|^2\cdot\frac{1}{[i(\epsilon_t-\epsilon_{L}(\vec{k}_{FWM}))+\gamma_L]}\\\nonumber
& &\frac{1}{[i(\epsilon_\tau-\epsilon_{L}(\vec{k}_2)-\epsilon_{U}(\vec{k}_2))+\gamma_L+\gamma_U]}.
\end{eqnarray}
Similarly,
\begin{eqnarray}\nonumber
&S^{(K)}&(\epsilon_\tau,\epsilon_t)\\\nonumber
&\propto&-|\Omega_{L}|^2|\Omega_{U}|^2\cdot\frac{1}{[i(\epsilon_t-\epsilon_{2L}(\vec{k}_{FWM})+\epsilon_{L}(\vec{k}_{FWM}))+\gamma_L]}\\\nonumber
& &\frac{1}{[i(\epsilon_\tau-\epsilon_{L}(\vec{k}_2)-\epsilon_{U}(\vec{k}_2))+\gamma_L+\gamma_U]},
\end{eqnarray}
\begin{eqnarray}\nonumber
S^{(L)}(\epsilon_\tau,\epsilon_t)
&\propto&|\Omega_{L}|^2|\Omega_{U}|^2\cdot\frac{1}{[i(\epsilon_t-\epsilon_{L}(\vec{k}_{FWM}))+\gamma_L]}\\\nonumber
& &\frac{1}{[i(\epsilon_\tau-\epsilon_{L}(\vec{k}_2)-\epsilon_{U}(\vec{k}_2))+\gamma_L+\gamma_U]},
\end{eqnarray}
and
\begin{eqnarray}\nonumber
&S^{(M)}&(\epsilon_\tau,\epsilon_t)\\\nonumber
&\propto&-|\Omega_{L}|^2|\Omega_{U}|^2\cdot\frac{1}{[i(\epsilon_t-\epsilon_{2L}(\vec{k}_{FWM})+\epsilon_{L}(\vec{k}_{FWM}))+\gamma_L]}\\\nonumber
& &\frac{1}{[i(\epsilon_\tau-\epsilon_{L}(\vec{k}_2)-\epsilon_{U}(\vec{k}_2))+\gamma_L+\gamma_U]}.
\end{eqnarray}
 Here, we do not repeat the same discussion for the LPUP-UP and the UP-UP groups. We plot the calculated 2DFT spectrum of the two-quantum contribution from all pathways in Fig. \ref{fig:sim_analytical} (b). Energy shifts, originating from self and cross-interactions $\epsilon_{2L}(\vec{k})\neq 2\epsilon_{L}(\vec{k})$ and $\epsilon_{LU}(\vec{k})\neq \epsilon_{L}(\vec{k})+\epsilon_{U}(\vec{k})$, are necessary for the appearance of the on and off-diagonal peaks respectively. Again, the diagonal and off-diagonal groups are associated with the self and cross-interactions respectively. Notice that the diagonal line in the figure ($\epsilon_\tau=2\epsilon_t$) is defined with an absorption energy that is twice the emission energy, this is characteristic of a ``two-quantum regime". In Fig. \ref{fig:sim_analytical} (b), the dashed diagonal line does not pass through the normal branches (NB) of 2LP-LP and 2UP-UP peaks. This is a consequence of the polaritons energy-momentum dispersion: $\epsilon_{L(U)}(\vec{k}_{FWM})=\epsilon_{L(U)}(\vec{k}_1)> \epsilon_{L(U)}(\vec{k}_2)=\epsilon_{L(U)}(0)$ (See Fig. 3(b)). 

\section{IV. THEORETICAL MODEL}
Our starting point is the bosonic exciton Hamiltonian including the exciton-exciton interaction and the exciton-photon coupling,
\begin{equation}
\hat{H}_{ex}=\hat{H}_{lin}+\hat{H}_{xint}.
\label{ecHamiltonian}
\end{equation}
The linear term $\hat{H}_{lin}$ and nonlinear exciton-exciton interaction term $\hat{H}_{xint}$ are respectively given by
\begin{eqnarray}
\hat{H}_{lin}&=&\int d{\bf r}\left[\hat{\bm \psi}_x^{\dagger}(\epsilon_x-\frac{\hbar^2\nabla^2}{2m_x})\hat{\bm \psi}_x+\hat{\bm \psi}_c^{\dagger}(\epsilon_c-\frac{\hbar^2\nabla^2}{2m_c})\hat{\bm \psi}_c\right.\nonumber\\
& &+\left.\frac{\Omega}{2} (\hat{\bm \psi}_x^{\dagger}\hat{\bm \psi}_c+\hat{\bm \psi}_c^{\dagger}\hat{\bm \psi}_x)+\frac{\Omega_{qm}}{2}(\hat{\bm \psi}_c^{\dagger}{\bm \psi}_b+{\bm \psi}_b^{*}\hat{\bm \psi}_c)\right]
\label{ecHamiltonian_lin}
\end{eqnarray}
and
\begin{equation}
\hat{H}_{xint}=\frac{1}{2}g_0\hat{\bm \psi}_x^{\dagger}\hat{\bm \psi}_x^{\dagger}\hat{\bm \psi}_x\hat{\bm \psi}_x.
\label{ecHamiltonian_int}
\end{equation}
$\hat{\bm \psi}_x$ ($\hat{\bm \psi}_x^{\dagger}$) and $\hat{\bm \psi}_c$ ($\hat{\bm \psi}_c^{\dagger}$) are the exciton and cavity photon annihilation (creation) operators respectively. The Rabi splitting between exciton and cavity photon is represented by $\Omega$. $\Omega_{qm}$ is a quasi-mode Rabi splitting which is the coupling of photons between the inside and outside of the cavity \cite{Ciuti2000}. ${\bm \psi}_b$ represents a classical photon field outside the cavity. $\epsilon_x$ and $\epsilon_c$ are respectively the exciton and photon eigenenergies. The exciton-exciton interaction constant is given by $g_0$ \cite{a}. Now, we introduce the polariton bases $\psi_{L}$ and $\psi_{U}$ defined as,
\begin{displaymath}
\left( \begin{array}{c}
\hat{\bm \psi}_x \\
\hat{\bm \psi}_c \\
\end{array} \right)
=
\left( \begin{array}{cc}
X & -C \\
C & \ X 
\end{array} \right)
\left( \begin{array}{c}
\hat{\bm \psi}_{L} \\
\hat{\bm \psi}_{U} \\
\end{array} \right),
\end{displaymath}
to rewrite the Hamiltonian Eq. \ref{ecHamiltonian_lin}. $X$ and $C$ are respectively excitonic (photonic) and photonic (excitonic) Hopfield coefficients of lower (upper) polaritons. They are chosen to be,
\begin{equation}\nonumber
X=\sqrt{\frac{1}{2} \left(1+\frac{\delta}{\sqrt{\delta^2+\Omega^2}}\right)}
\end{equation}
and
\begin{equation}\nonumber
\displaystyle
C=-\sqrt{\frac{1}{2} \left(1-\frac{\delta}{\sqrt{\delta^2+\Omega^2}}\right)},
\end{equation}
which diagonalizes a non-interacting exciton-photon Hamiltonian at k=0. We rewrite $\hat{H}_{lin}$ in terms of polariton basis under a parabolic approximation of polariton energy-momentum dispersion: $\epsilon_{\rm L(U)}(\vec{k})\simeq\epsilon_{L,0}+\frac{\hbar^2}{2m_{L(U)}}\vec{k}^2$. The Hamiltonian of the linear part in polariton basis reads,
\begin{eqnarray}
\hat{H}_{lin}&\simeq &\hat{H}'_{lin}\nonumber \\
&=&\int d{\bf r}\left[ \hat{\bm \psi}_{L}^{\dagger}(\epsilon_{L,0}-\frac{\hbar^2\nabla^2}{2m_{L}})\hat{\bm \psi}_{L}+\hat{\bm \psi}_{U}^{\dagger}(\epsilon_{U,0}-\frac{\hbar^2\nabla^2}{2m_{U}})\hat{\bm \psi}_{U}\right.\nonumber \\
& &+\frac{\Omega_{L}^*}{2}\hat{\bm \psi}_{L}^{\dagger}{\bm \psi}_b+\frac{\Omega_{L}}{2}{\bm \psi}_b^{*}\hat{\bm \psi}_{L}\nonumber \\
& &+\left.\frac{\Omega_{U}^*}{2}\hat{\bm \psi}_{U}^{\dagger}{\bm \psi}_b+\frac{\Omega_{U}}{2}{\bm \psi}_b^{*}\hat{\bm \psi}_{U}\right].
\label{lupHamiltonian_lin}
\end{eqnarray} 
$\epsilon_{L,0}$ and $\epsilon_{U,0}$ are respectively the energies of lower and upper-polariton at zero momentum written as,
\begin{equation}\nonumber
\epsilon_{L,0}=\frac{1}{2}\left(2\epsilon_x+\delta-\sqrt{\delta^2+\Omega^2}\right)
\end{equation}
and
\begin{equation}\nonumber
\epsilon_{U,0}=\frac{1}{2}\left(2\epsilon_x+\delta+\sqrt{\delta^2+\Omega^2}\right).
\end{equation}
The polariton quasi-mode Rabi splittings are defined as $\Omega_L=C\Omega_{qm}$ and $\Omega_U=X\Omega_{qm}$. The polariton mass is given by $1/m_{L(U)}=|X|^2/m_{x(c)}+|C|^2/m_{c(x)}$. Now, the linear Hamiltonian Eq. \ref{lupHamiltonian_lin} is formally the same as that of two oscillators, which are weakly coupled to photonic fields. Actually, this is why the double-sided Feynman diagram analysis, which is normally used in a weak-coupling system between oscillators and photons, can be applied to this system in which exciton and photon are strongly coupled inside a microcavity. We introduce polariton-polariton interactions as,
\begin{eqnarray}
\hat{H}_{pint}&=&\frac{1}{2}g_L\hat{\bm \psi}_{L}^{\dagger}\hat{\bm \psi}_{L}^{\dagger}\hat{\bm \psi}_{L}\hat{\bm \psi}_{L}+\frac{1}{2}g_U\hat{\bm \psi}_{U}^{\dagger}\hat{\bm \psi}_{U}^{\dagger}\hat{\bm \psi}_{U}\hat{\bm \psi}_{U}.\nonumber \\
& &+g_X\hat{\bm \psi}_{L}^{\dagger}\hat{\bm \psi}_{L}\hat{\bm \psi}_{U}^{\dagger}\hat{\bm \psi}_{U}.
\label{lupHamiltonian_int}
\end{eqnarray} 
This is a simple effective Hamiltonian that can give self- and cross-interactions between lower and upper-polaritons. Using the Hopfield coefficients, we set $g_{L} = g_0|X|^4$ , $g_{U} = g_0|C|^4$, and $g_X =2g_0|X|^2|C|^2$. The relation between this Hamiltonian and exciton-exciton interaction Hamiltonian ($\hat{H}_{xint}$) is discussed in Appendix C. Only when the kinetic term (energy-momentum dispersion) in Eq. \ref{ecHamiltonian_lin} is neglected, can we derive the effective polariton-polariton Hamiltonian ($\hat{H}_{pint}$) from exciton-exciton interaction Hamiltonian ($\hat{H}_{xint}$) using a perturbation theory in the low density regime. 

\begin{figure}[h]
\includegraphics[width=0.35\textwidth]{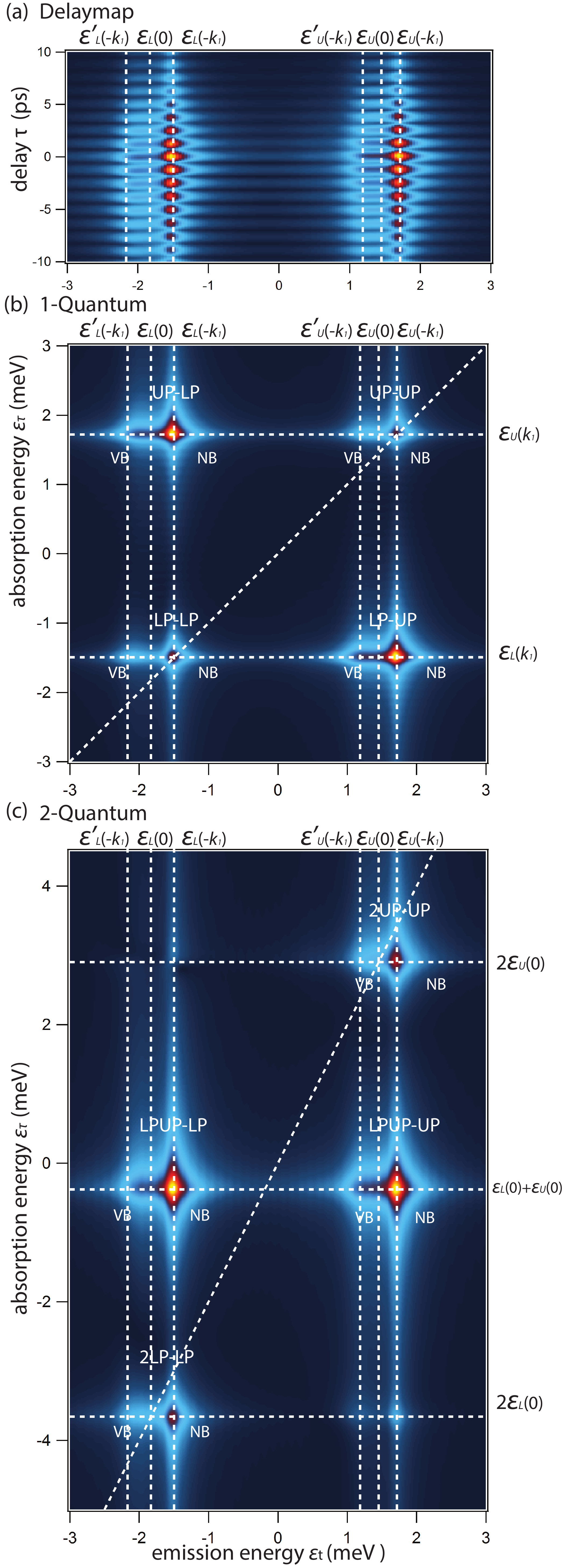}
\caption{Simulated amplitude of FWM spectrum as a function of emission energy and $k_2$-$k_1$ pulse delay $\tau$: $|S(\tau,\epsilon_t)|$ (a). Simulated amplitude of 2D FWM spectrum $|S(\epsilon_\tau,\epsilon_t)|$ for the (b) 1-quantum and (c) 2-quantum regions.}
\label{fig:sim_PRB}
\end{figure}

In order to understand the experimental observations in detail, we perform non-perturbative numerical simulations. A polariton basis total Hamiltonian is,
\begin{equation}
\hat{H}_{pol}=\hat{H}'_{lin}+\hat{H}_{pint}.
\end{equation}
With the aid of the Heisenberg equations of motion $i\hbar\frac{d}{dt}{\hat{\bm \psi}}_{L(U)}=[\hat{\bm \psi}_{L(U)},H_{pol}]$ and the mean-field approximation \cite{Carusotto2013}, the equations of motion of lower and upper polariton wavefunctions are simplified to non-equilibrium lower-upper polariton Gross-Pitaevskii equations:
\begin{eqnarray}
i\hbar \dot{\psi}_{L}&=&(\epsilon_{L,0}-\frac{\hbar^2}{2m_L}\nabla^2+g_{L}|\psi_{L}|^2\nonumber \\
& &+g_{X}|\psi_{U}|^2-i\frac{\gamma_L}{2})\psi_{L}+\frac{\Omega_L^*}{2}f_{\rm ext}\\
i\hbar \dot{\psi}_{U}&=&(\epsilon_{U,0}-\frac{\hbar^2}{2m_U}\nabla^2+g_{U}|\psi_{U}|^2\nonumber \\
& &+g_{X}|\psi_{L}|^2-i\frac{\gamma_{U}}{2})\psi_{U}+\frac{\Omega_U^*}{2}f_{\rm ext},
\label{lpupGP}
\end{eqnarray}
where $\psi_{L(U)}=\langle \hat{\psi}_{L(U)}\rangle$ is lower (upper) polariton wave function. The polariton decay rate is given by $\gamma_{L(U)}= |X|^2\gamma_{x(c)}+|C|^2\gamma_{c(x)}$, where $\gamma_{x}$ and $\gamma_{c}$ are chosen to be the same (0.33 meV). The $k_2$ and $k_1$ pulse excitations are represented by an external photon field $f_{\rm ext}(={\bm \psi}_b)$. The constant $g_{L(U)}$ represents a self-interaction of lower (upper) polaritons, while $g_{X}$ is a cross-interaction constant between the lower and upper-polariton. In analogy with nonlinear optics, $g_{L(U)}$ and $g_{X}$ can be called the self-phase modulation (SPM) and the cross-phase modulation (XPM) terms respectively. Similar to nonlinear optics, the XPM term $g_X$ is twice as strong as the SPM term $g_{L(U)}$ \cite{Agrawal2007}. In the simulation, the exciton-exciton interaction constant $g_0$ is set as 2 meV/$n_0$, where $n_0$ is a normalized density.  The excitation $\frac{1}{2}\Omega_{qm}^{*}f_{\rm ext}$ is a Gaussian pulse with a peak intensity of $0.5\sqrt{n_0}$ and a pulse duration of 250 fs. This model describes coherent processes and includes mean field interaction. Using this simplified model, we can directly investigate the contribution of self and cross-polariton interactions through the three nonlinear interaction constants: $g_{L}$, $g_{U}$, and $g_X$. This is a key advantage of this model over the exciton-photon basis (local mode basis) Gross-Pitaevskii equations (See Appendix C). In the local mode basis, the only nonlinear interaction constant represents the exciton-exciton interaction, thus we cannot deal with self and cross-interactions of lower and upper polaritons independently. Simulated FWM spectra are presented in Fig. \ref{fig:sim_PRB}. All numerical simulations are performed in one-dimensional space. Similar to experimental observations (Fig. \ref{fig:exp_PRB} (b) and (c)), the simulated 2DFT spectra (Fig. \ref{fig:sim_PRB} (b) and (c)) clearly show fine structures inside the four peak groups. 

\subsection{A. Different interaction contributions}
\begin{figure}[h]
\includegraphics[width=0.49\textwidth]{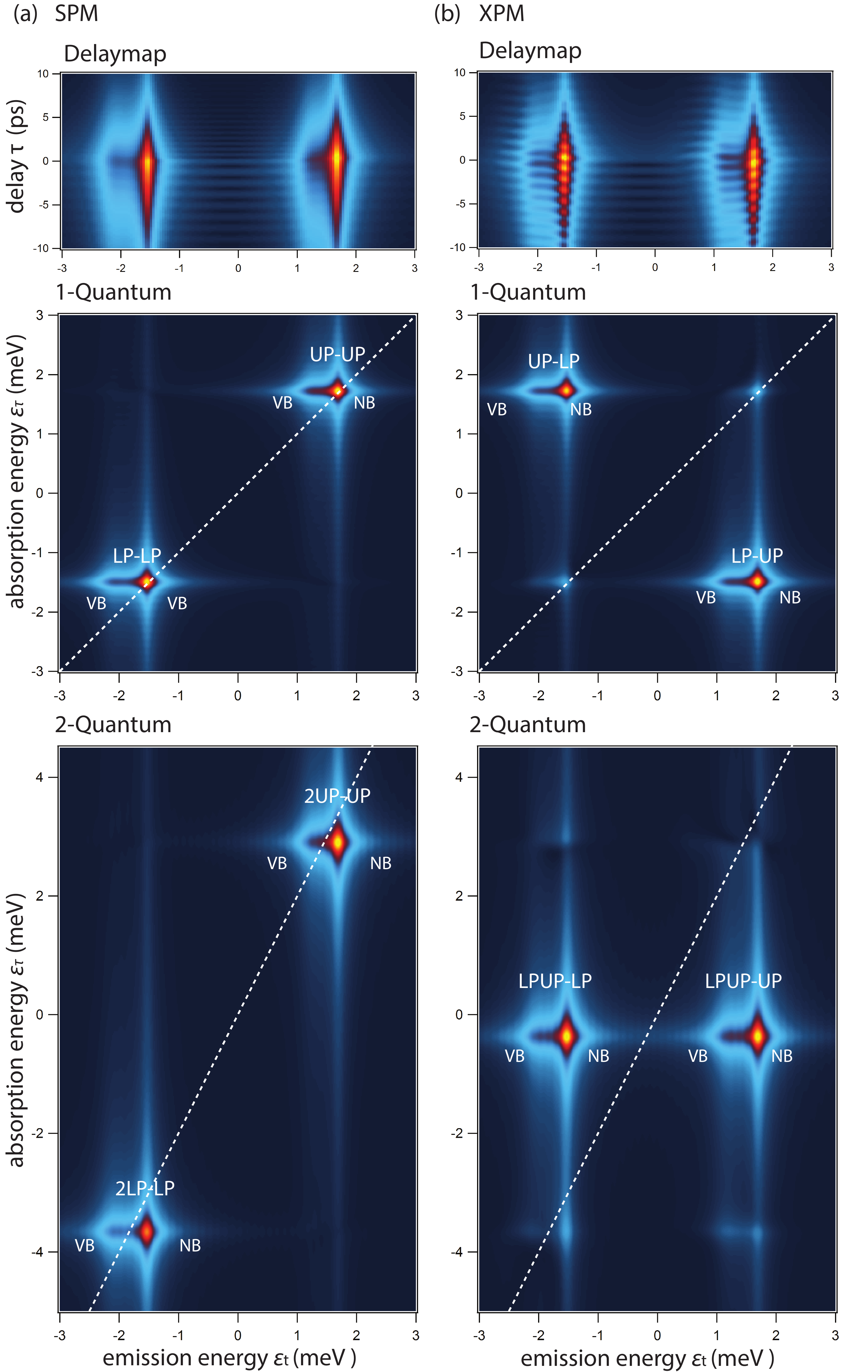}
\caption{Simulated delaymap $|S(\tau,\epsilon_t)|$ and 2D FT spectrum $|S(\epsilon_\tau,\epsilon_t)|$ for 1-quantum and 2-quantum regime. Simulations are performed including only self-interaction (or self phase modulation: SPM) $g_{X}=0$ (a) and for only cross-interaction (or cross phase modulation: XPM) $g_{L(U)}=0$ (b).}
\label{fig:sim_SPM_XPM}
\end{figure}

In order to obtain better insight into the importance of the interactions, two different sets of spectra are calculated and plotted in Fig. \ref{fig:sim_SPM_XPM}. Firstly, Fig. \ref{fig:sim_SPM_XPM} (a) shows the 2D spectra when considering only self-interactions $g_{L(U)}$. As expected, we find two main LP-LP (2LP-LP) and UP-UP (2UP-UP) groups along the diagonal line in one (two)-quantum 2DFT spectrum. The LP-LP (2LP-LP) and UP-UP (2UP-UP) groups originate from lower-lower and upper-upper self-interactions respectively. On the other hand, the 2D spectra, including only the lower-upper cross-interaction $g_{X}$ (Fig. \ref{fig:sim_SPM_XPM} (b)), shows only the off-diagonal groups UP-LP (LPUP-LP) and LP-UP (LPUP-UP) in one(two)-quantum 2D spectrum. Only when we include both the self $g_{L(U)}$ and cross $g_X$ interactions, is the observed experimental 2D spectra are reproduced (Fig. \ref{fig:sim_PRB}). Notice that, since the cross-interaction constant $g_X$ is twice as strong as the self-interaction constant $g_{L(U)}$, around zero cavity detuning, the off-diagonal peaks are brighter than diagonal ones, corroborating the experimental results.

\subsection{B. Fine structures inside peak groups} 
\begin{figure}[H]
\begin{center}
\includegraphics[width=0.28\textwidth]{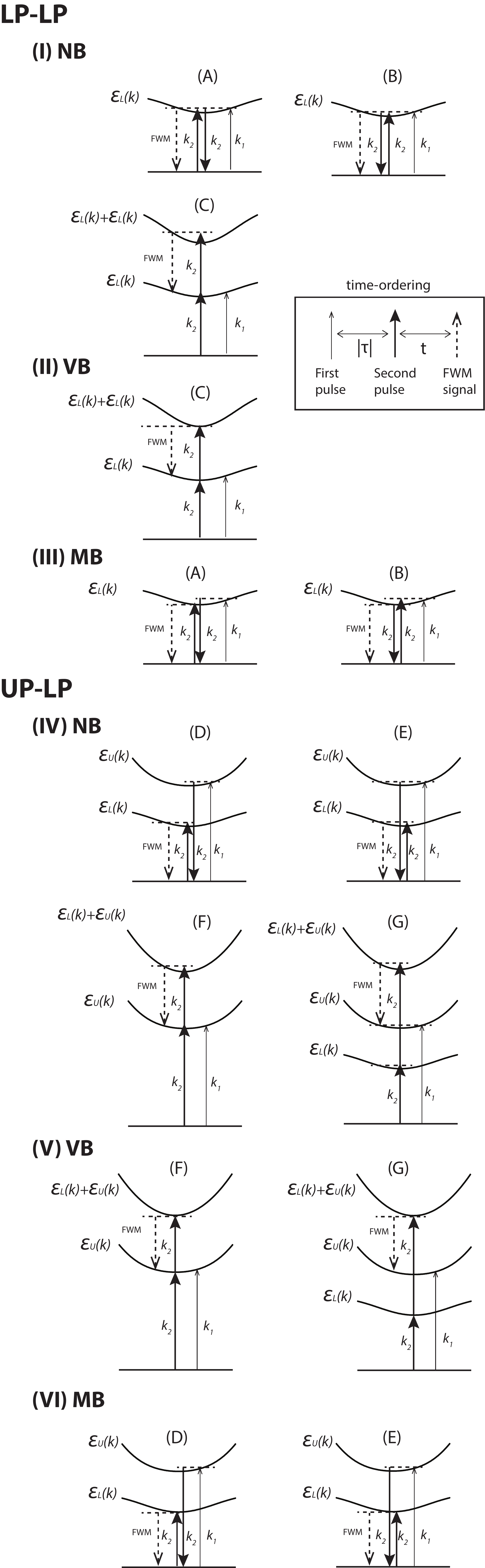}
\caption{Schematic energy diagrams representing the origin of the fine structures in one-quantum 2D spectrum. Three solid arrows represent two degenerate $k_2$ pulses and $k_1$ pulse. The dashed arrows are FWM emissions. For NB schematics, the indices (A)-(G) correspond to those of the double-sided Feynman diagrams in Fig. \ref{fig:negative_doublesided_PRB}}
\label{fig:energy_diagrams1}
\end{center}
\end{figure}

\begin{figure}[H]
\begin{center}
\includegraphics[width=0.27\textwidth]{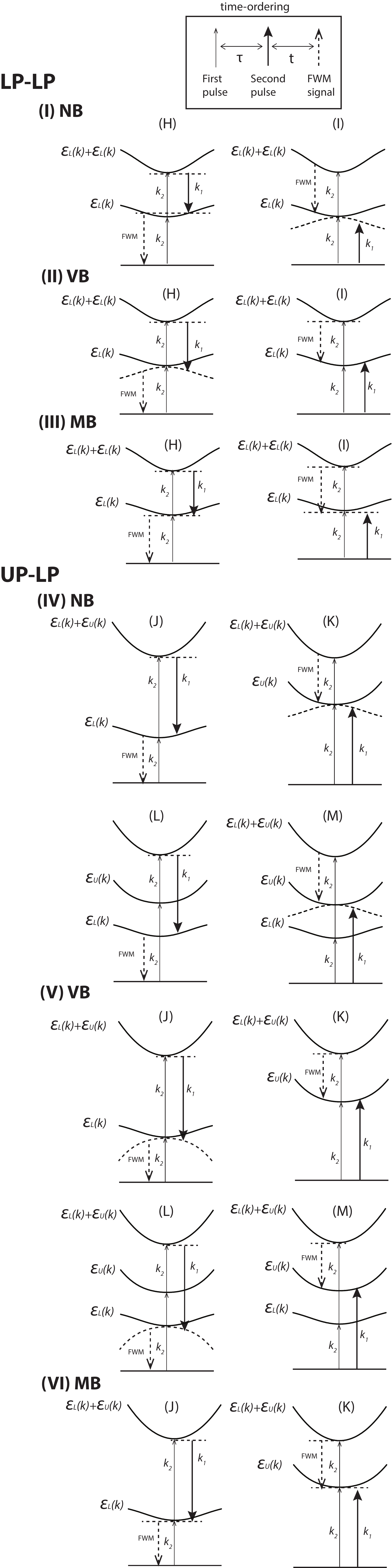}
\caption{Same schematic as Fig. \ref{fig:energy_diagrams1} for 2-quantum 2D spectrum. For normal branch (NB) schematics, the indices (H)-(M) correspond to those of the double-sided Feynman diagrams in Fig. \ref{fig:positive_doublesided_PRB}}
\label{fig:energy_diagrams2}
\end{center}
\end{figure}

In the 2DFT spectra, we can find a fine structure (the normal (NB), middle (MB), and virtual branch (VB)) inside each peak group. They can be classified by the emission energies. For example, in the LP-LP group, the emission energies of NB, MB and VB respectively correspond to $\epsilon_{L,0}+\frac{\hbar^2}{2m_L}k_1^2$, $\epsilon_{L,0}$ and $\epsilon_{L,0}-\frac{\hbar^2}{2m_L}k_1^2$. As mentioned above, the fine structure is related to the polaritons energy-momentum dispersion, associated with a light polariton mass. The idea is schematically shown in Fig. \ref{fig:energy_diagrams1} (one-quantum regime) and \ref{fig:energy_diagrams2} (two-quantum regime) as energy diagrams. In this section, we focus on the fine structure in the one-quantum regime however, the discussion is completely the same for the two-quantum regime. 

Firstly, we discuss the process associated with the normal branches. As is shown in Fig. \ref{fig:energy_diagrams1}(I) and (IV), NB emission is an on-branch FWM emission. Therefore, its emission energy is $\epsilon_{L(U)}(k_{FWM})=\epsilon_{L(U)}(-k_1)=\epsilon_{L(U),0}+\frac{\hbar^2}{2m_{L(U)}}k_1^2$ for both diagonal and off-diagonal peak groups. In the third-order perturbative calculation of FWM signal in Section III, only this branch appears (See Fig. \ref{fig:sim_analytical}), because the FWM signal evolve with the eigenenergy of polaritons, which corresponds to $\epsilon_{L(U)}(-k_1)$ (See Appendix B).  

Secondly, let us consider the virtual branches. The appearance of the VB can be understood in a framework of off-resonant scattering induced by polariton-polariton interactions \cite{Savvidis2001}. The fine structure results from momentum and energy conservation in the FWM process. As the energy diagrams of the VB, in Fig. \ref{fig:energy_diagrams1}. (II) and (V) show, if the $k_1$ and $k_2$ pulses are on-resonant excitation, energy and momentum conservation demand VB emission. Using energy conservation, for LP-LP group (See Fig. 8 (II)), the energy of VB emission is given by $\epsilon'_{L}(k_{FWM})=2\epsilon_{L}(\vec{k_2})-\epsilon_{L}(\vec{k_1})=\epsilon_{L,0}-\frac{\hbar^2}{2m_L}k_1^2$. Since NB emission energy is $\epsilon_{L,0}+\frac{\hbar^2}{2m_L}k_1^2$, the separation between NB and VB emission is $2\frac{\hbar^2}{m_L}k_1^2$, which is twice the energy difference between the $k_2$ and $k_1$ transmission peak of the lower polariton. In the experiment, the energy separation between NB and VB emission is 0.58 meV (See Fig. \ref{fig:exp_PRB} (b) and (c)). This energy difference corresponds to the twice of 0.29 meV, which is the energy difference of the $k_1$ and $k_2$ transmission peak of the lower polariton presented in Fig. \ref{fig:schematic} (c). Additionally, this relation holds for both for negative and positive cavity detuning (see Sect. IV C). Therefore, we can safely say that VB peak in LP-LP group originates from this process. The VB energy-momentum dispersion $\epsilon'_{L}(k)$ is a mirror image of NB energy-momentum dispersion $\epsilon_{L}(k)$. We note that in the weak intensity excitation regime, the dispersion is still parabolic and we do not consider linearisation of the dispersion due to the Bogoliubov transformation \cite{Kohnle2011}. In UP-LP group (See Fig. \ref{fig:energy_diagrams1} (V)), similar to the LP-LP group, from momentum and energy conservation VB emission energy is $\epsilon'_{LU}(k_{FWM})=\epsilon_{L}(\vec{k_2})+\epsilon_{U}(\vec{k_2})-\epsilon_{U}(\vec{k_1})=\epsilon_{L,0}-\frac{\hbar^2}{2m_U}k_1^2$. Notice that, in Fig. \ref{fig:energy_diagrams1} (V), the higher energy-momentum dispersion is the sum of two different energy-momentum dispersions $\epsilon_{L}(\vec{k})$ and $\epsilon_{U}(k)$. While the VB of the UP-LP group is visible in the simulation, Fig. \ref{fig:sim_PRB}, experimentally we cannot find the VB in the UP-LP group (See Fig. \ref{fig:exp_PRB}). The reason for this is not clear. We can apply the same discussion to the LP-UP and the UP-UP groups, but experimentally the VB cannot be found in the 2DFT spectra of these groups.

Now, let us focus on the middle branches. In the UP-LP group of the experimental 2DFT spectra, a strong peak is found next to NB. The emission energy of this peak corresponds to $\epsilon_{L}(\vec{k_2})=\epsilon_{L,0}$, which is located in the middle of NB and VB. Thus we name this peak MB. In Fig. \ref{fig:exp_PRB}, we can identify a weak MB in LP-LP. Additionally, the experimental peak in UP-UP is also considered as a MB because the emission energy of the UP-UP peak is $\epsilon_{U}(\vec{k_2})=\epsilon_{U,0}$. On the other hand, the numerically simulated 2DFT spectra does not include MBs, which makes the interpretation of MB difficult. Finally, schematically, we associated the MB to the processes described in Fig. \ref{fig:energy_diagrams1}(III) and (VI). These are the processes where energy and  momentum conservation are satisfied and one of the degenerate fields of the second pulse is not resonant to the energy-momentum dispersion. To understand the detailed mechanism of the MB, a further investigation and model are necessary. For instance, taking into account effects such as, excitation induced dephasing (EID) and relaxation of the upper-polariton into exciton reservoirs. 

\subsection{C. FWM spectra at $\delta$=-2 meV and  $\delta$=2.7 meV}
\begin{figure}[H]
\includegraphics[width=0.48\textwidth]{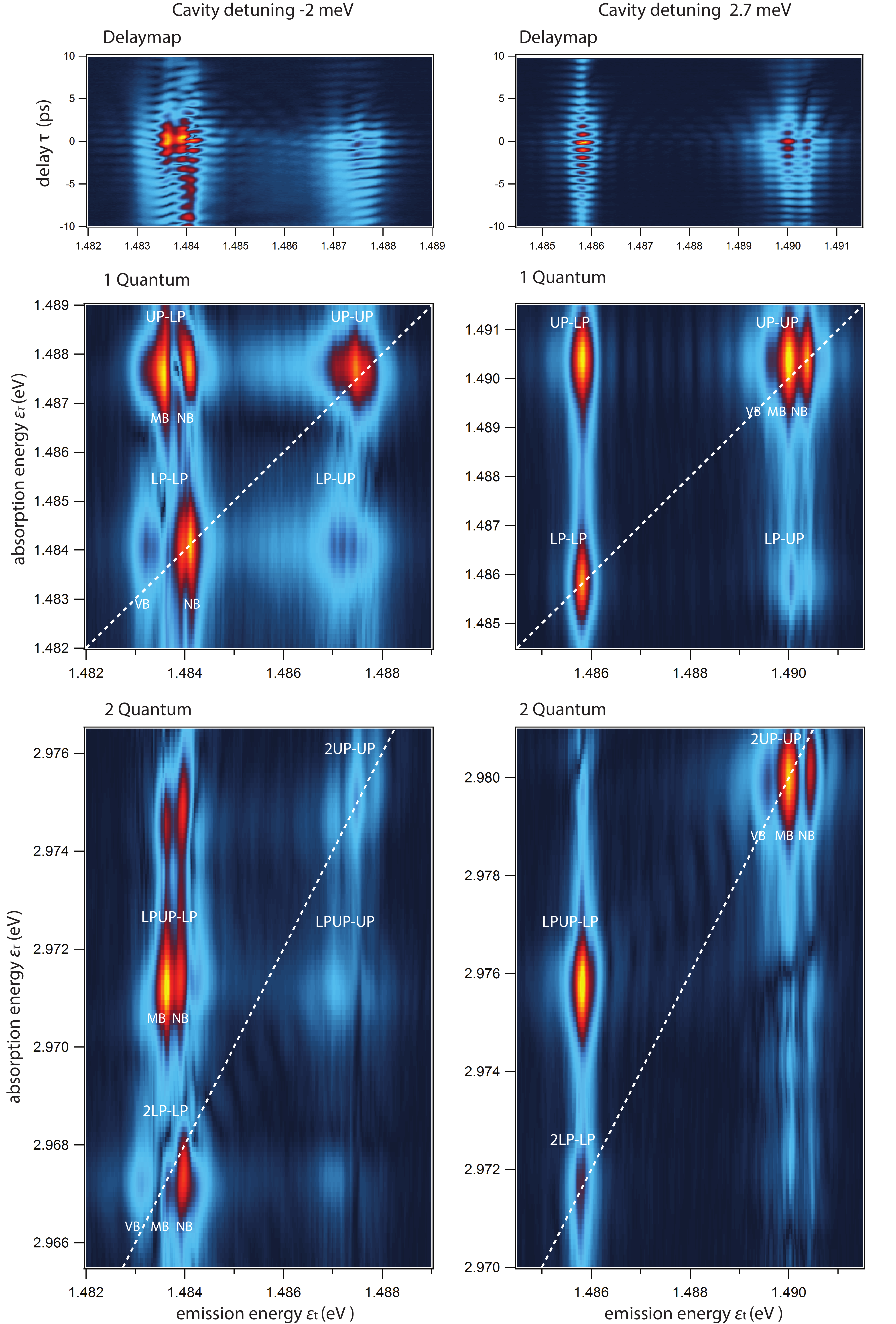}
\caption{Experimental FWM delaymap $|S(\tau,\epsilon_t)|$ and 1- and 2-quantum 2D FT spectrum $|S(\epsilon_\tau,\epsilon_t)|$ for negative ($\delta=-2$ meV) and for positive ($\delta=2.7$ meV) cavity detuning.}
\label{fig:exp_negative_positive_PRB}
\end{figure}

To obtain more insight into the origin of the fine structure energy, we discuss the results obtained at both negative $\delta$=-2.0 meV and positive $\delta$=2.7 meV cavity detuning. In Fig. \ref{fig:exp_negative_positive_PRB}, we display the amplitude  of the 2DFT signal $|S(\epsilon_\tau,\epsilon_t)|$. We observe the same features presented in Fig. \ref{fig:exp_PRB} (b) and (c), however the emission energy separation between the NB and the VB varies depending on the cavity detuning due to the change in the polaritons energy-momentum dispersion. 
By detuning the cavity photon energy far below the exciton energy $\delta$=-2.0 meV, the LP becomes photon-like and acquires a lighter mass than the UP. Due to the lighter mass of the LP at $\delta$=-2.0 meV, the NB and VB in the LP-LP group are more separated than in the experimental results at the cavity detuning $\delta=-0.38$ meV (Fig. \ref{fig:exp_PRB} (a)). 

On the other hand, positive detuning $\delta=2.7$ meV results in a flatter energy-momentum dispersion for the LP. The resulting heavier mass makes the fine structures in the LP-LP group almost degenerate. We also find three fine structure peaks in UP-UP (2UP-UP) group. This is due to the lighter mass of the UP with positive cavity detuning. However, the brightest MB peak cannot be described within our simple model. It might be related to dynamic redshift of the UP mean field energy \cite{Kohnle2011,Kohnle2012}.  We also observe weaker spectral features in the one- and two-quantum 2D spectra, which could originate from higher order non-linear processes that were not completely eliminated in our heterodyne detection scheme. This will be the subject of further investigations to confirm their origin. 

\begin{figure}[h]
\includegraphics[width=0.45\textwidth]{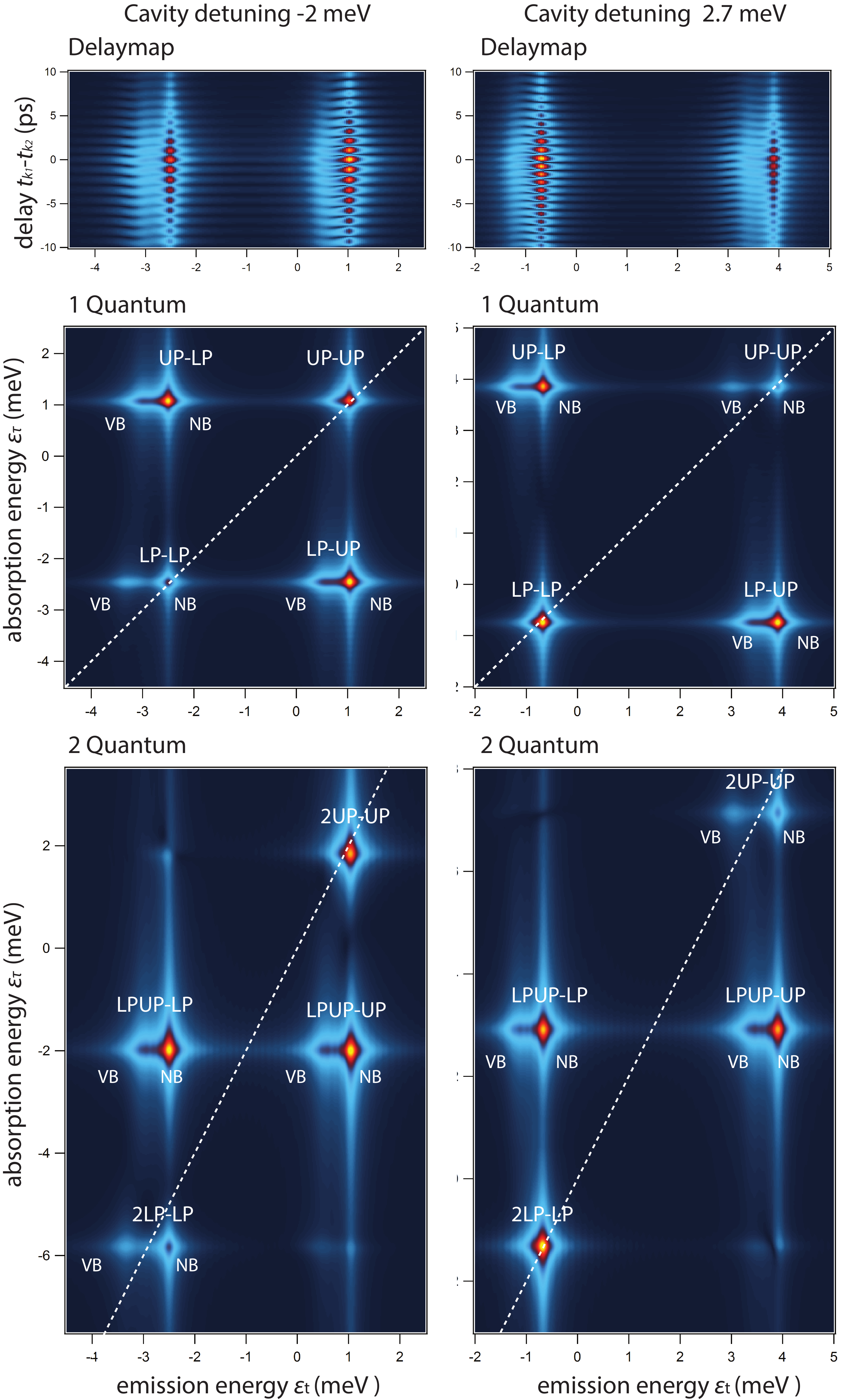}
\caption{Simulated FWM delaymap $|S(\tau,\epsilon_t)|$ and 1 and 2-quantum 2D FT spectrum $|S(\epsilon_\tau,\epsilon_t)|$ for negative ($\delta=-2$ meV) and for positive ($\delta=2.7$ meV) cavity detuning.}
\label{fig:sim_n2meV_p2meV}
\end{figure}

The simulations for negative ($\delta=-2$ meV) and positive ($\delta=2.7$ meV) cavity detuning are shown in Fig. \ref{fig:sim_n2meV_p2meV}. As expected, for the negative detuning the energy separation of fine structures (NB and VB) in the LP-LP group increases because the LP dispersion becomes photon-like. Conversely, for positive detuning, we cannot distinguish the fine structure inside the LP-LP group due to the flat dispersion of the LP (exciton-like). The fine structures in UP-UP group shows the inverse detuning dependence as the LP-LP group: a small separation for $\delta=-2$ meV and a large separation for $\delta=2.7$ meV. Notice that the energy separations between NB and VB in UP-LP (LPUP-LP) and LP-UP (LPUP-UP) groups are the average of those of LP-LP (2LP-LP) and UP-UP (2LP-LP) groups. 

\subsection{D. Comparison with experiments}
Our numerical model qualitatively explains the appearance of LP-LP (2LP-LP) and UPLP (LPUP-LP) peak groups and the fine structures inside them. In principle, as is shown in the simulations, the fine structure should appear in the group LP-UP (2LP-UP) in the experimental spectrum. However, experimentally there is a strong amplitude asymmetry between UP-LP (LPUP-LP) and LP-UP (LPUP-UP) group. Compared with UP-LP (LPUP-LP) group, the LP-UP (LPUP-UP) group is very weak and we cannot resolve detailed structures. This type of asymmetry in the off-diagonal peaks is observed also in Ref. \cite{1367-2630-15-2-025005} and in 2DFT spectra of bare quantum well \cite{Yang2007,Nardin2014} experiments. Additionally, the fine structure in UP-UP group is more complicated than what is computed within our simple model. The theoretically predicted off-diagonal VBs do not appear in the experiment. On the other hand, MBs, which are not visible in the simulation, appear in the experiment. 

\begin{figure}[h]
\includegraphics[width=0.35\textwidth]{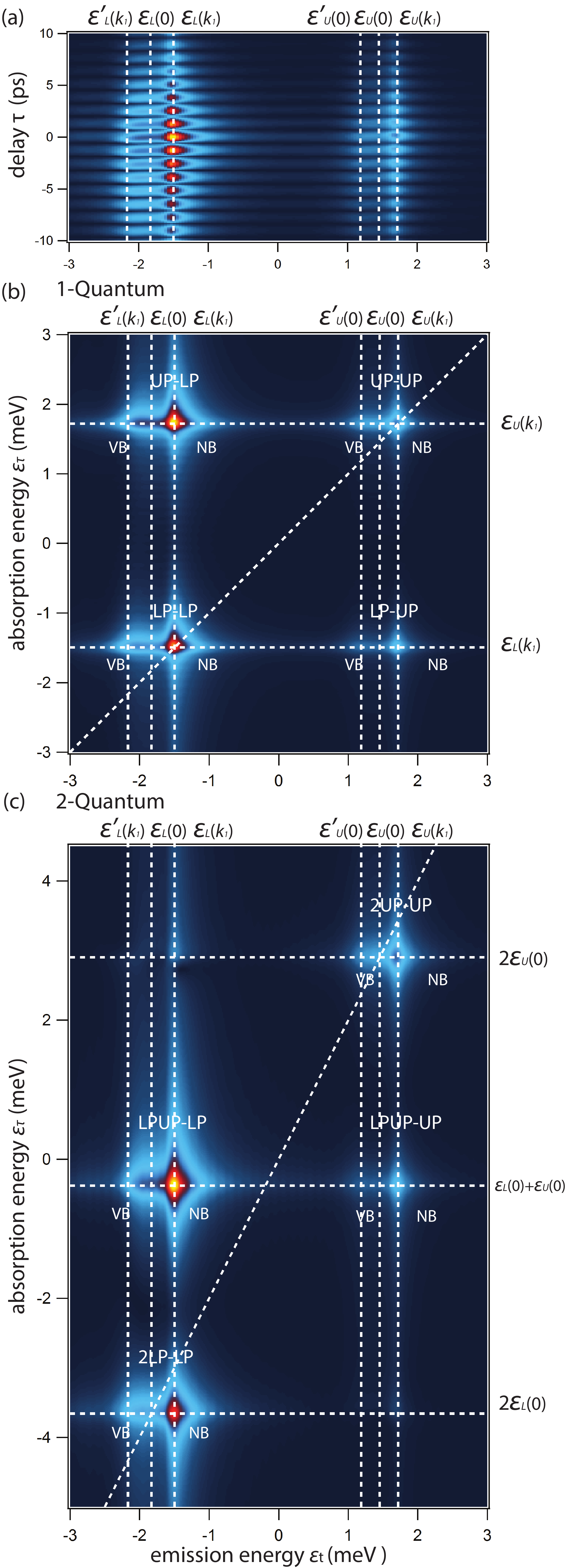}
\caption{Simulation based on interaction constants as free parameters. The cavity detuning is $\delta=-0.38$ meV. The interaction constants set as $g'_{L}:g'_{U}:g'_{LU}:g'_{UL}=1:0.6:0.3:1.6$ (meV/$n_0$). Amplitude of FWM spectrum as a function of emission energy and $k_2$-$k_1$ pulse delay $\tau$: $|S(\tau,\epsilon_t)|$ (a). Simulated amplitude of 2D FWM spectrum $|S(\epsilon_\tau,\epsilon_t)|$ for the (b) 1-quantum and (c) 2-quantum regions.}
\label{fig:sim_asymmetric}
\end{figure}

Although a microscopic model completely explaining the experiments is still lacking, in term of asymmetric amplitudes of off-diagonal peaks, we can attempt to fit the observed spectra by dealing with the polariton Gross-Pitaevskii equations in a more phenomenological way. Here, for the fitting between the experiment of the cavity detuning $\delta=-0.38$ meV and simulation, we consider the strength of the interaction constants of polaritons as free fitting parameters:
\begin{eqnarray}
i\hbar \dot{\psi}_{L}&=&(\epsilon_{L,0}-\frac{\hbar^2}{2m_L}\nabla^2+g'_{L}|\psi_{L}|^2\nonumber \\
& &+g'_{UL}|\psi_{U}|^2-i\frac{\gamma_L}{2})\psi_{L}+\frac{\Omega_L^*}{2}f_{\rm ext}\\
i\hbar \dot{\psi}_{U}&=&(\epsilon_{U,0}-\frac{\hbar^2}{2m_U}\nabla^2+g'_{U}|\psi_{U}|^2\nonumber \\
& &+g'_{LU}|\psi_{L}|^2-i\frac{\gamma_{U}}{2})\psi_{U}+\frac{\Omega_U^*}{2}f_{\rm ext}.
\end{eqnarray}
The above equations are formally same as Eq. \ref{lpupGP}. However, the interaction constants ($g'_L$, $g'_U$, $g'_{LU}$, and $g'_{UL}$) are fitting parameters and no more connected to the exciton-exciton interaction Hamiltonian $\hat{H}_{int}$. This means that $g'_{LU}$ is not necessary equal to $g'_{UL}$. This is the advantage of using lower- and upper-polariton basis equations compared with a conventional exciton-photon Gross-Pitaevskii equations, where the amplitude of LP-UP group is always same as that of UP-LP groups. We set the interaction constants as $g'_{L}:g'_{U}:g'_{LU}:g'_{UL}=1:0.6:0.3:1.6$ (meV/$n_0$), which correspond to the ratio of the integrated amplitudes of peaks groups LP-LP, UP-UP, LP-UP and UP-LP in Fig. \ref{fig:exp_PRB}(b). The other parameters are the same as those in the previous section. The simulated 2DFT spectra with these parameters are shown in Fig. \ref{fig:sim_asymmetric}. We find that our phenomenological model reproduces more closely the asymmetric intensities of off-diagonal peaks. The relative strength of the interaction constants is believed to reflect neglected contributions such as a frequency dependent non-Markovian nature of exciton-exciton interaction (exciton-exciton correlation) \cite{Oestreich1995,Savasta2003,1367-2630-15-2-025005} and photon-assisted exchange interaction between polaritons that reinforces the repulsive interaction among the lower polaritons but that weakens the repulsive interaction among the upper polaritons \cite{Combescot2007,a}.

\section{V. CONCLUSION}
In summary, 2D Fourier transformation spectroscopy is performed to investigate polariton-polariton interactions in semiconductor microcavities.
The experimental 2D optical spectra demonstrate the existence of lower-upper polaritons cross-interaction and of lower (upper) polariton self-interaction, which originate from the Coulomb and exchange interactions between the fermion constituents of the exciton-polariton. Furthermore, an asymmetry of the coupling between the upper and lower polaritons is clearly evidenced in these spectra and indicates complex many-body effects such as exciton-exciton correlation and photon-assisted exchange scattering between carriers constituting the exciton-polaritons. In addition, a fine structure in the emission energy is identified as resulting from the polariton energy-momentum dispersion and the optical non-linearity of the third order. This work opens the way for a quantitative study of many-body effects on composite bosons based on two-dimensional Fourier transform spectroscopy.

\section{ACKNOWLEDGEMENT}
The present work is supported by the Swiss National Science Foundation under project N$^{\circ}$135003, the Quantum Photonics National Center of Competence in research N$^{\circ}$115509 and the European Research Council under project Polaritonics contract N$^{\circ}$291120. The polatom network is also acknowledged.
%\bibliography{biblio}

\section{APPENDIX A: NUMERICAL PHASE CORRECTION}
Due to the lack of phase stabilization in our experimental setup, we apply a numerical phase correction process to the 2D spectrum \cite{Kasprzak2011,Albert2013}. Phase corrected 2D spectrum $S_{\rm cor}(\epsilon_t,\epsilon_\tau )$ can be obtained through the Fourier transformation of $S_{\rm cor}(\epsilon_t,\tau )$, where $S_{\rm cor}(\epsilon_t,\tau )=S(\epsilon_t,\tau )\exp\bigl(\frac{i}{\hbar}(\epsilon_{\rm cor}\tau-\arg [S(\epsilon_{\rm cor},\tau)])\bigr)$. Here $\epsilon_{\rm cor}$ represents a phase correction frequency. In the article, we chose the upper polariton energy as the phase correction energy. In the case of Fig. \ref{fig:exp_PRB}, where cavity detuning is $\delta=-0.38$, $\epsilon_{\rm cor}$ is equal to 1.4883eV.

\begin{figure}[h]
\includegraphics[width=0.4\textwidth]{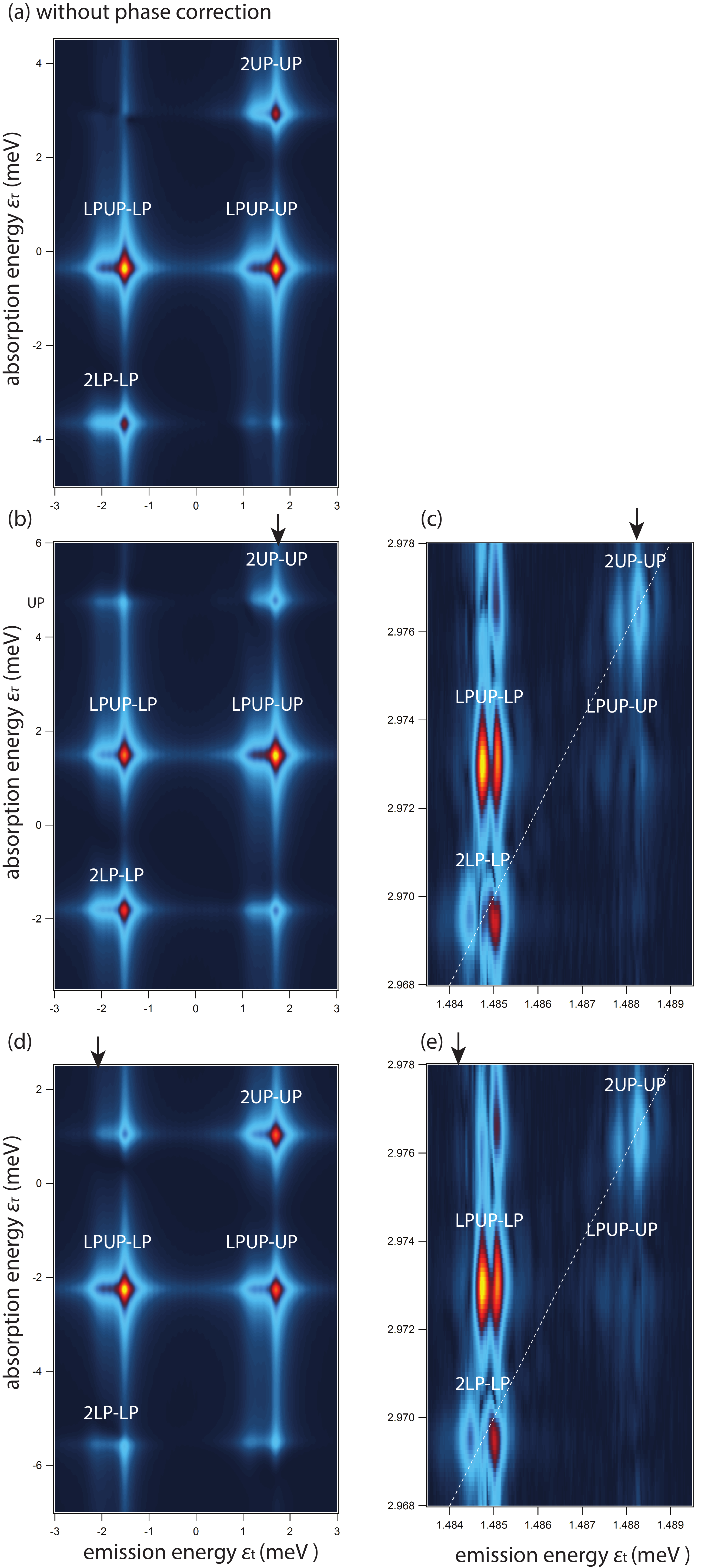}
\caption{Simulated two-quantum 2DFT spectra $|S(\epsilon_\tau,\epsilon_t)|$ without phase correction (a), which is same spectrum used in Fig. \ref{fig:exp_PRB} (c). Simulated (b,d) and experimental (c,e) two-quantum 2DFT spectra $|S(\epsilon_\tau,\epsilon_t)|$ for choosing upper-polariton (b-c) and lower-polariton (d-e) as phase correction energies. Arrows represent the phase correction energies $\epsilon_{\rm cor}$.}
\label{fig:phasecorrection}
\end{figure}

In Fig. \ref{fig:phasecorrection}, we present experimental and simulated two-quantum 2D spectra with different phase correction energies. Fig. \ref{fig:phasecorrection} (b-c) and (d-e) are respectively two-quantum 2DFT spectra using the NB in upper-polariton and VB in lower-polariton as phase correction energies. Both simulated and experimental spectra indicate that phase correction process mainly shifts the $\epsilon_\tau$ axis and affects the amplitude of 2D spectrum, but it does not change the fine structures of the peaks.

\section{Appendix B: Third-order perturbation theory}
In this appendix, we briefly explain the calculation of the FWM signal based on third-order perturbation theory. Firstly, we introduce an ideal three-levels  exciton system interacting with classical electric fields. $|g\rangle$ represents a ground state. A first $|e\rangle$ and second excited states $|m\rangle$ are respectively one-exciton and two-excitons states. Now, a density matrix can be constructed as following:
\begin{equation}
\rho
=
\left( \begin{array}{ccc}
\rho_{gg} & \rho_{eg}^* & \rho_{mg}^*\\
\rho_{eg} & \rho_{ee} & \rho_{me}^*\\
\rho_{mg} & \rho_{me} & \rho_{mm}\\
\end{array} \right)
\end{equation}
The system's Hamiltonian $H$ is composed of an eigen Hamiltonian $H_0$ and an exciton-electric field coupling $\mu$: $H=H_0+\mu$. $H_0$ and $\mu$ are given by, 
\begin{equation}
H_0
=
\left( \begin{array}{ccc}
\epsilon_g & 0 & 0\\
0 & \epsilon_e & 0\\
0 & 0 & \epsilon_m\\
\end{array} \right)
\end{equation} 
and
\begin{equation}
\mu
=
\left( \begin{array}{ccc}
0 & \Omega E^* & 0\\
\Omega^* E & 0 & \sqrt{2}\Omega E^*\\
0 & \sqrt{2}\Omega^* E & 0\\
\end{array} \right).
\end{equation} 
$\epsilon_g$, $\epsilon_e$, and $\epsilon_m$ are respectively the ground, first, and second excited state energies. The time evolution of the density matrix $\rho$ is determined by the Liouville-von Neumann equation $i\hbar \dot{\rho}=[H,\rho]=H\rho-\rho H$. Firstly, without considering the momentum degree of freedom, we can obtain 6 coupled equations of motion for the density matrix elements. Then, we perturbatively expand the density matrix in terms of the orders of the incident electric fields as   
\begin{equation}
\rho_{ij}=\rho_{ij}^{(0)}+\rho_{ij}^{(1)}+\rho_{ij}^{(2)}+\rho_{ij}^{(3)}+...
\end{equation} 
The first, second and third-order density matrices are given by
\begin{eqnarray}
i\hbar \dot{\rho}_{eg}^{(1)} &=& (\epsilon_e-\epsilon_g-i\gamma)\rho_{eg}^{(1)}+\Omega ^* E\rho_{gg}^{(0)}\nonumber\\
\nonumber\\
i\hbar \dot{\rho}_{gg}^{(2)} &=& -(\Omega ^*E\rho_{eg}^{(1)*}-\Omega E^*\rho_{eg}^{(1)})\nonumber\\
i\hbar \dot{\rho}_{ee}^{(2)} &=& -i\Gamma_e\rho_{ee}+ (\Omega ^*E\rho_{eg}^{(1)*}-\Omega E^*\rho_{eg}^{(1)})\nonumber\\
i\hbar \dot{\rho}_{mg}^{(2)} &=& (\epsilon_m-\epsilon_g-i\gamma)\rho_{mg}^{(2)}+\sqrt{2}\Omega^* E\rho_{eg}^{(1)}\label{eq:third-order}\\
\nonumber\\
i\hbar \dot{\rho}_{eg}^{(3)} &=& (\epsilon_e-\epsilon_g-i\gamma)\rho_{eg}^{(3)}\nonumber\\
& &+\Omega ^* E(\rho_{gg}^{(2)}-\rho_{ee}^{(2)})+\sqrt{2}\Omega E^*\rho_{mg}^{(2)}\nonumber\\
i\hbar \dot{\rho}_{me}^{(3)} &=& (\epsilon_m-\epsilon_e-i\gamma)\rho_{me}^{(3)}+\sqrt{2}\Omega^* E\rho_{ee}^{(2)}-\Omega  E^* \rho_{mg}^{(2)}.\nonumber
\end{eqnarray}
$\Gamma_{e}$ and $\gamma$ respectively represent phenomenological decay rates of the population of the first excited state and polarization. Now we introduce the degree of freedom of momentum and extend the density matrix into 9$\times$9 taking into account three different momentum channels: $\vec{k}_2(=0)$, $\vec{k}_1$, and $\vec{k}_{FWM}(=-\vec{k}_1)$. We refer to the three channels respectively as ``pump", ``trigger", and ``idler". Since a superposition state between two different momentum channels does not appear within the following third-order perturbation calculation, the 9$\times$9 density matrix is block diagonalized into three 3$\times$3 matrices (pump, trigger, and idler). We substitute the density matrix,  
\begin{equation}
\rho_{ij}^{(n)}=\rho_{ij,p}^{(n)}+\rho_{ij,t}^{(n)}e^{i\vec{k}_1\cdot\vec{x}}+\rho_{ij,i}^{(n)}e^{-i\vec{k}_1\cdot\vec{x}}
\end{equation} 
and electric fields,
\begin{equation}
E=E_p(t)+E_t(t)e^{i\vec{k}_1\vec{x}}.
\end{equation} 
into Eq. \ref{eq:third-order} and select combinations that satisfy momentum conservation (phase-matching condition). The two pulses $E_p$ and $E_t$ respectively correspond to $\vec{k}_2$ and $\vec{k}_1$ pulses in the main text. These combinations differ between the one-quantum (negative delay) and two-quantum regime (positive delay). Firstly, in the one-quantum regime, there are three pathways A, B, and C: 
\begin{eqnarray}
{\rm path\ A}\nonumber\\
i\hbar \dot{\rho}_{eg,t}^{(1)} &=& (\epsilon_e^t-\epsilon_g^t-i\gamma)\rho_{eg,t}^{(1)}+\Omega ^* E_t\rho_{gg,p}^{(0)}\nonumber\\
i\hbar \dot{\rho}_{gg,i}^{(2)} &=& -\Omega ^* E_p\rho_{eg,t}^{(1)*}\label{dif_pathA}\\
i\hbar \dot{\rho}_{eg,i}^{(3),A} &=& (\epsilon_e^i-\epsilon_g^i-i\gamma)\rho_{eg,i}^{(3),A}+\Omega ^* E_p\rho_{gg,i}^{(2)}\nonumber
\end{eqnarray}
\begin{eqnarray}
{\rm path\ B}\nonumber\\
i\hbar \dot{\rho}_{eg,t}^{(1)} &=& (\epsilon_e^t-\epsilon_g^t-i\gamma)\rho_{eg,t}^{(1)}+\Omega ^* E_t\rho_{gg,p}^{(0)}\nonumber\\
i\hbar \dot{\rho}_{ee,i}^{(2)} &=& -i\Gamma_e\rho_{ee,i}^{(2)}+\Omega ^* E_p\rho_{eg,t}^{(1)*}\label{dif_pathB}\\
i\hbar \dot{\rho}_{eg,i}^{(3),B} &=& (\epsilon_e^i-\epsilon_g^i-i\gamma)\rho_{eg,i}^{(3),B}-\Omega ^* E_p\rho_{ee,i}^{(2)}\nonumber
\end{eqnarray}
\begin{eqnarray}
{\rm path\ C}\nonumber\\
i\hbar \dot{\rho}_{eg,t}^{(1)} &=& (\epsilon_e^t-\epsilon_g^t-i\gamma)\rho_{eg,t}^{(1)}+\Omega ^* E_t\rho_{gg,p}^{(0)}\nonumber\\
i\hbar \dot{\rho}_{ee,i}^{(2)} &=& -i\Gamma_e\rho_{ee,i}^{(2)}+\Omega ^* E_p\rho_{eg,t}^{(1)*}\label{dif_pathC}\\
i\hbar \dot{\rho}_{me,i}^{(3),C} &=& (\epsilon_m^i-\epsilon_e^i-i\gamma)\rho_{me,i}^{(3),C}+\sqrt{2}\Omega^* E_p\rho_{ee,i}^{(2)}\nonumber
\end{eqnarray}
Here, $\epsilon_{e(m)}^p$, $\epsilon_{e(m)}^t$, and $\epsilon_{e(m)}^i$ respectively represent eigen energies of the pump, trigger and idler momentum: $\epsilon_{e(m)}^p=\epsilon_{e(m)}(\vec{k}_2)$, $\epsilon_{e(m)}^t=\epsilon_{e(m)}(\vec{k}_1)$, and $\epsilon_{e(m)}^i=\epsilon_{e(m)}(\vec{k}_{FWM})$. Secondly, in the two-quantum regime (positive delay), two pathways H and I exists:
\begin{eqnarray}
{\rm path\ H}\nonumber\\
i\hbar \dot{\rho}_{eg,p}^{(1)} &=& (\epsilon_e^p-\epsilon_g^p-i\gamma)\rho_{eg,p}^{(1)}+\Omega ^* E_p\rho_{gg,p}^{(0)}\nonumber\\
i\hbar \dot{\rho}_{mg,p}^{(2)} &=& (\epsilon_m^p-\epsilon_g^p-i\gamma)\rho_{mg,p}^{(2)}+\sqrt{2}\Omega^* E_p\rho_{eg,p}^{(1)}\label{dif_pathH}\\
i\hbar \dot{\rho}_{eg,i}^{(3),H} &=& (\epsilon_e^i-\epsilon_g^i-i\gamma)\rho_{eg,i}^{(3),H}+\sqrt{2}\Omega E^*_t\rho_{mg,p}^{(2)}\nonumber
\end{eqnarray}
\begin{eqnarray}
{\rm path\ I}\nonumber\\
i\hbar \dot{\rho}_{eg,p}^{(1)} &=& (\epsilon_e^p-\epsilon_g^p-i\gamma)\rho_{eg,p}^{(1)}+\Omega ^* E_p\rho_{gg,p}^{(0)}\nonumber\\
i\hbar \dot{\rho}_{mg,p}^{(2)} &=& (\epsilon_m^p-\epsilon_g^p-i\gamma)\rho_{mg,p}^{(2)}+\sqrt{2}\Omega^* E_p\rho_{eg,p}^{(1)}\label{dif_pathI}\\
i\hbar \dot{\rho}_{me,i}^{(3),I} &=& (\epsilon_m^i-\epsilon_e^i-i\gamma)\rho_{me,i}^{(3),I}-\Omega  E^*_t\rho_{mg,p}^{(2)}\nonumber
\end{eqnarray}
The labels A-I correspond to those in the main text. These pathways are represented by the double-sided Feynman diagrams presented in Fig. \ref{fig:negative_doublesided_PRB} and \ref{fig:positive_doublesided_PRB}. Finally, with the third-order polarization density matrices, the third-order polarization $P^{(3)}$ is calculated as 
\begin{eqnarray}
P^{(3)}+c.c. &=& Tr[\rho^{(3)}_i\mu]\nonumber\\
&=& \Omega \rho_{eg,i}^{(3)}+\sqrt{2}\Omega\rho_{me,i}^{(3)}+c.c.,
\label{pol_sum1}
\end{eqnarray}
Now, we directly integrate Eq. \ref{dif_pathA}-\ref{dif_pathI} and calculate the third-order polarization density matrices. In general, the two pulses are written as 
\begin{equation}
E_p(t)=\widetilde{E}_p(t)e^{-(i/\hbar)\epsilon_{pu}(t-t_p)}
\end{equation}
and
\begin{equation}
E_t(t)=\widetilde{E}_t(t)e^{-(i/\hbar)\epsilon_{tr}(t-t_t)}.
\end{equation}
$\widetilde{E}_p(t)$ and $\widetilde{E}_t(t)$ are respectively pulse envelopes. $\epsilon_{pu(tr)}$ and $t_{p(t)}$ are the center energy and arrival time of the pulse. For an analytical integration, we assume the two pulses have delta function  envelopes:
\begin{equation}
\widetilde{E}_{p}(t)=\widetilde{E}^0_{p}\delta (t-t_p)\ \ {\rm and}\ \ \widetilde{E}_{t}(t)=\widetilde{E}^0_{t}\delta (t-t_t).
\end{equation}
Let us consider one-quantum regime and the path A. We can directly integrate Eq. \ref{dif_pathA} as following,
\begin{eqnarray*}
\rho_{eg,t}^{(1)}(t) &=& \frac{1}{i\hbar}\Omega^*  e^{-(i/\hbar)(\epsilon_e^t-\epsilon_g^t-i\gamma)t}e^{(i/\hbar)\epsilon_{tr}t_t}\\
& &\cdot\int_{-\infty}^tdt' \widetilde{E}_t(t')e^{(i/\hbar)(\epsilon_e^t-\epsilon_g^t-i\gamma-\epsilon_{tr})t'}\rho_{gg,p}^{(0)}\\
&=& \frac{1}{i\hbar}\Omega^*  \widetilde{E}^0_te^{-(i/\hbar)(\epsilon_e^t-\epsilon_g^t-i\gamma)(t-t_t)}\rho_{gg,p}^{(0)}\\
\\
\rho_{gg,i}^{(2)}(t) &=& -(\frac{1}{i\hbar})\Omega^* e^{(i/\hbar)\epsilon_{pu}t_p}\int_{-\infty}^tdt' \widetilde{E}_p(t')e^{(i/\hbar)(-\epsilon_{pu})t'}\rho_{eg,t}^{(1)*}(t')\\
&=& -(\frac{1}{i\hbar})\Omega^*  \widetilde{E}^0_p\ \rho_{eg,t}^{(1)*}(t_p)\\
&=& (\frac{1}{i\hbar})^2|\Omega |^2\widetilde{E}^0_p\widetilde{E}_t^{0*}e^{(i/\hbar)(\epsilon_e^t-\epsilon_g^t+i\gamma)(t_p-t_t)}\rho_{gg,p}^{(0)*}\\
\\
\rho_{eg,i}^{(3),A}(t) &=& (\frac{1}{i\hbar})\Omega^*  e^{-(i/\hbar)(\epsilon_e^i-\epsilon_g^i-i\gamma)t}e^{(i/\hbar)\epsilon_{pu}t_p}\\
& &\cdot\int_{-\infty}^tdt' \widetilde{E}_p(t')e^{(i/\hbar)(\epsilon_e^i-\epsilon_g^i-i\gamma-\epsilon_{pu})t'}\rho_{gg,i}^{(2)}(t')\\
&=& (\frac{1}{i\hbar})\Omega^* \widetilde{E}^0_pe^{-(i/\hbar)(\epsilon_e^i-\epsilon_g^i-i\gamma)(t-t_p)}\rho_{gg,i}^{(2)}(t_p)\\
&=& (\frac{1}{i\hbar})^3\Omega^* |\Omega |^2\widetilde{E}_p^{0}\widetilde{E}_p^{0}\widetilde{E}_t^*\\
& &\cdot e^{-(i/\hbar)(\epsilon_e^i-\epsilon_g^i-i\gamma)(t-t_p)}e^{(i/\hbar)(\epsilon_e^t-\epsilon_g^t+i\gamma)(t_p-t_t)}\rho_{gg,p}^{(0)*}.
\end{eqnarray*}
Redefining the times as $t-t_p\rightarrow t$ and $t_t-t_p\rightarrow\tau$ and recalling Eq. \ref{pol_sum1}, the signal contributing from the path A is given by
\begin{equation*}
S^{(3),A}(t,\tau)\propto |\Omega |^4e^{-(i/\hbar)(\epsilon_e^i-\epsilon_g^i-i\gamma)t}e^{(i/\hbar)(\epsilon_e^t-\epsilon_g^t+i\gamma)|\tau|}
\end{equation*}
Similarly, the signal associated with the path B and C are calculated as  
\begin{equation*}
S^{(3),B}(t,\tau)\propto |\Omega |^4e^{-(i/\hbar)(\epsilon_e^i-\epsilon_g^i-i\gamma)t}e^{(i/\hbar)(\epsilon_e^t-\epsilon_g^t+i\gamma)|\tau|}
\end{equation*}
and
\begin{equation*}
S^{(3),C}(t,\tau)\propto -2|\Omega |^4e^{-(i/\hbar)(\epsilon_m^i-\epsilon_e^i-i\gamma)t}e^{(i/\hbar)(\epsilon_e^t-\epsilon_g^t+i\gamma)|\tau|}
\end{equation*}
In two-quantum regime (the positive delay), there are two pathways H and I. We directly integrate Eq. \ref{dif_pathH} and obtain the third-order polarization in the following way. 
\begin{eqnarray*}
\rho_{eg,p}^{(1)}(t) &=& \frac{1}{i\hbar}\Omega ^* e^{-(i/\hbar)(\epsilon_e^p-\epsilon_g^p-i\gamma)t}e^{(i/\hbar)\epsilon_{pu}t_p}\\
& &\cdot\int_{-\infty}^tdt' \widetilde{E}_p(t')e^{(i/\hbar)(\epsilon_e^p-\epsilon_g^p-i\gamma-\epsilon_{pu})t'}\rho_{gg,p}^{(0)}\\
&=& \frac{1}{i\hbar}\Omega ^* \widetilde{E}_p^0e^{-(i/\hbar)(\epsilon_e^p-\epsilon_g^p-i\gamma)(t-t_p)}\rho_{gg,p}^{(0)}\\
\\
\rho_{mg,p}^{(2)}(t) &=& \frac{1}{i\hbar}\sqrt{2}\Omega^* e^{-(i/\hbar)(\epsilon_m^p-\epsilon_g^p-i\gamma)t}e^{(i/\hbar)\epsilon_{pu}t_p}\\
& &\cdot\int_{-\infty}^tdt' \widetilde{E}_p(t')e^{(i/\hbar)(\epsilon_m^p-\epsilon_g^p-i\gamma-\epsilon_{pu})t'}\rho_{eg,p}^{(1)}(t')\\
&=& \frac{1}{i\hbar}\sqrt{2}\Omega^*\widetilde{E}^0_p e^{-(i/\hbar)(\epsilon_m^p-\epsilon_g^p-i\gamma)(t-t_p)}\rho_{eg,p}^{(1)}(t_p)\\
&=& (\frac{1}{i\hbar})^2\sqrt{2}\Omega^*\Omega ^* \widetilde{E}_p^0\widetilde{E}_p^0\ e^{-(i/\hbar)(\epsilon_m^p-\epsilon_g^p-i\gamma)(t-t_p)}\rho_{gg,p}^{(0)}\\
\\
\rho_{eg,i}^{(3),D}(t) &=& (\frac{1}{i\hbar})\sqrt{2}\Omega e^{-(i/\hbar)(\epsilon_e^i-\epsilon_g^i-i\gamma)t}e^{(i/\hbar)\epsilon_{tr}t_t}\\
& &\cdot\int_{-\infty}^tdt' \widetilde{E}_t^*(t')e^{(i/\hbar)(\epsilon_e^i-\epsilon_g^i-i\gamma-\epsilon_{tr})t'}\rho_{mg,p}^{(2)}(t')\\
&=& (\frac{1}{i\hbar})\sqrt{2}\Omega\widetilde{E}_t^{0*} e^{-(i/\hbar)(\epsilon_e^i-\epsilon_g^i-i\gamma)(t-t_t)}\rho_{mg,p}^{(2)}(t_t)\\
&=& (\frac{1}{i\hbar})^32|\Omega|^2\Omega ^*\widetilde{E}_t^{0*}\widetilde{E}_p^0\widetilde{E}_p^0e^{-(i/\hbar)(\epsilon_e^i-\epsilon_g^i-i\gamma)(t-t_t)}\\
& &\cdot e^{-(i/\hbar)(\epsilon_m^p-\epsilon_g^p-i\gamma)(t_t-t_p)}\rho_{gg,p}^{(0)}.\\
\end{eqnarray*}
With a redefinition the times, $t-t_t\rightarrow t$ and $t_t-t_p\rightarrow|\tau|$ and Eq. \ref{pol_sum1}, the signal contributing from the path H is given by
\begin{equation*}
S^{(3),H}(t,\tau)\propto 2|\Omega |^4e^{-(i/\hbar)(\epsilon_e^i-\epsilon_g^i-i\gamma)t}e^{-(i/\hbar)(\epsilon_m^p-\epsilon_g^p-i\gamma)|\tau|}.
\end{equation*}
In the same way, the signal associated with the path I is calculated as   
\begin{equation*}
S^{(3),I}(t,\tau)\propto -2|\Omega |^4e^{-(i/\hbar)(\epsilon_m^i-\epsilon_e^i-i\gamma)t}e^{-(i/\hbar)(\epsilon_m^p-\epsilon_g^p-i\gamma)|\tau|}.
\end{equation*}
With the aid of the double-sided Feynman diagrams, it is not difficult to extend the present discussion to the case where two different exciton modes are coupled. Note that we can easily apply our calculation to the polariton system in the main text just by rewriting the state $|e\rangle$ and $|m\rangle$ respectively to $|L\rangle$ ($|U\rangle$) and $|2L\rangle$ ($|2U\rangle$). In this case , the ground state energy is set to be zero and the following replacement holds: $\epsilon_e^p\rightarrow\epsilon_{L(U)}(\vec{k}_2)$, $\epsilon_m^p\rightarrow\epsilon_{2L(2U)}(\vec{k}_2)$, $\epsilon_e^t\rightarrow\epsilon_{L(U)}(\vec{k}_1)$, $\epsilon_m^t\rightarrow\epsilon_{2L(2U)}(\vec{k}_1)$, $\epsilon_e^i\rightarrow\epsilon_{L(U)}(-\vec{k}_{FWM})$ and $\epsilon_m^i\rightarrow\epsilon_{2L(2U)}(-\vec{k}_{FWM})$. The coupling between polaritons and classical electric fields outside a cavity is represented by the quasi-mode coupling $\frac{1}{2}\Omega_{L(U)}$. 
\section{Appendix C: SIMULATION BASED ON EXCITON-PHOTON BASIS (LOCAL MODE SIMULATION)}
In this appendix, we discuss the connection between exciton-exciton interaction Hamiltonian Eq. \ref{ecHamiltonian_int} and the polariton-polariton interaction Eq. \ref{lupHamiltonian_int}. Here, we assume that there is no motional (kinetic) degree of freedom by neglecting the energy momentum dispersion (the kinetic term) of exciton and photon. Firstly, in the exciton-photon basis, the linear term $\hat{H}_{lin}$ and exciton-exciton interaction term $\hat{H}_{int}$ are
\begin{equation}
\hat{H}_{lin}=\epsilon_x\hat{\psi}_x^{\dagger}\hat{\psi}_x+\epsilon_c\hat{\psi}_c^{\dagger}\hat{\psi}_c+\frac{\Omega}{2} (\hat{\psi}_x^{\dagger}\hat{\psi}_c+\hat{\psi}_c^{\dagger}\hat{\psi}_x)
\end{equation} 
and 
\begin{equation}
\hat{H}_{int}=\frac{1}{2}g_0\hat{\psi}_{x}^{\dagger}\hat{\psi}_{x}^{\dagger}\hat{\psi}_{x}\hat{\psi}_{x}.
\end{equation} 
In the polariton basis, we can diagonalize the linear term $\hat{H}_{0}$ as
\begin{equation}
\hat{H}_{lin}=\epsilon_{L,0}\hat{\psi}_{L}^{\dagger}\hat{\psi}_{L}+\epsilon_{U,0}\hat{\psi}_{U}^{\dagger}\hat{\psi}_{U}
\end{equation} 
The exact expression of the exciton-exciton interaction term $\hat{H}_{int}$ in polariton basis is the following: 
\begin{align}
\hat{H}_{int}&=\frac{1}{2}g_0|X|^4\hat{\psi}_{L}^{\dagger}\hat{\psi}_{L}^{\dagger}\hat{\psi}_{L}\hat{\psi}_{L}\tag{a1}\\
& +\frac{1}{2}g_0|C|^4\hat{\psi}_{U}^{\dagger}\hat{\psi}_{U}^{\dagger}\hat{\psi}_{U}\hat{\psi}_{U}\tag{a2}\\
& +2g_0|X|^2|C|^2\hat{\psi}_{L}^{\dagger}\hat{\psi}_{U}^{\dagger}\hat{\psi}_{L}\hat{\psi}_{U}\tag{a3}\\
& -g_0|X|^2X^*C\hat{\psi}_{L}^{\dagger}\hat{\psi}_{L}^{\dagger}\hat{\psi}_{L}\hat{\psi}_{U}\tag{b1}\\
& -g_0|C|^2X^*C\hat{\psi}_{L}^{\dagger}\hat{\psi}_{U}^{\dagger}\hat{\psi}_{U}\hat{\psi}_{U}\tag{b2}\\
& -g_0|X|^2XC^*\hat{\psi}_{U}^{\dagger}\hat{\psi}_{L}^{\dagger}\hat{\psi}_{L}\hat{\psi}_{L}\tag{b3}\\
& -g_0|C|^2XC^*\hat{\psi}_{U}^{\dagger}\hat{\psi}_{U}^{\dagger}\hat{\psi}_{U}\hat{\psi}_{L}\tag{b4}\\
& +\frac{1}{2}g_0X^{*2}C^2\hat{\psi}_{L}^{\dagger}\hat{\psi}_{L}^{\dagger}\hat{\psi}_{U}\hat{\psi}_{U}\tag{c1}\\
& +\frac{1}{2}g_0C^{*2}X^2\hat{\psi}_{U}^{\dagger}\hat{\psi}_{U}^{\dagger}\hat{\psi}_{L}\hat{\psi}_{L}.\tag{c2}
\end{align}
(a1)-(a3) are respectively lower and upper polaritons self and cross-interactions employed in the main text. The terms (b1)-(b4) couple lower and upper-polaritons depending on the number of polaritons. The last terms (c1) and (c2) annihilate two upper (lower) polaritons and create two lower (upper) polaritons. These terms are called the Darling-Dennison coupling terms \cite{Hamm2011}. Now we will evaluate the energy of the Hamiltonian $\hat{H}=\hat{H}_{lin}+\hat{H}_{int}$ using a standard stationary perturbation theory of quantum mechanics \cite{Cohen-Tannoudji1977}. The linear term $\hat{H}_{lin}$ is a unperturbated Hamiltonian and the eigenstate of $\hat{H}_{lin}$ is written as $|n,m\rangle$, where $n$ and $m$ are respectively the number of lower and upper-polaritons. $\hat{H}_{int}$ is the perturbative Hamiltonian. The energy $E_{n,m}=\langle n,m|\hat{H}|n,m\rangle$ is evaluated as  
\begin{eqnarray}
E_{n,m}&=&E^0_{n,m}+\langle n,m|\hat{H}_{int}|n,m\rangle\nonumber\\
& &+\sum_{(n',m')\neq(n,m)}\frac{|\langle n',m'|\hat{H}_{int}|n,m\rangle|^2}{E^0_{n,m}-E^0_{n',m'}}+...
\end{eqnarray}
The first term, $E^0_{n,m}$, is an unperturbed energy defined as $E^0_{n,m}=\langle n,m|\hat{H}_{0}|n,m\rangle=\epsilon_{L,0}n+\epsilon_{U,0}m$. The second and third terms respectively represent a first and second-order perturbation of the energy correction. A simple calculation shows that the self (a1)-(a2) and cross-interaction terms (a3) have diagonal elements and contribute to the first-order perturbation of energy. The other terms (b1)-(c2) contribute only to the second-order of the perturbation. This means that the self and cross-interaction terms can be considered as dominant terms in the perturbative regime. The explicit form of the energy is calculated as
\begin{align}
E_{n,m}&=\epsilon_{L,0}n+\epsilon_{U,0}m\nonumber\\
& +\frac{1}{2}g_0|X|^4n(n-1)\tag{a'1}\\
& +\frac{1}{2}g_0|C|^4m(m-1)\tag{a'2}\\
& +2g_0|X|^2|C|^2n\cdot m\tag{a'3}\\
& +\frac{g_0^2}{\Omega}|X|^6X|C|^2n^2(n+1)m\tag{b'1}\\
& +\frac{g_0^2}{\Omega}|C|^6X|C|^2(n+1)(m-1)^2m\tag{b'2}\\
& -\frac{g_0^2}{\Omega}|X|^6|C|^2(m+1)(n-1)^2n\tag{b'3}\\
& -\frac{g_0^2}{\Omega}|C|^6|X|^2(m+1)m^2n\tag{a'4}\\
& +\frac{g_0^2}{2\Omega}|X|^4|C|^4(n+2)(n+1)(m-1)m\tag{c'1}\\
& -\frac{g_0^2}{2\Omega}|C|^4|X|^4(m+2)(m+1)(n-1)n\tag{c'2}\\
& +...\nonumber
\end{align}
\begin{figure}[h]
\begin{center}
\includegraphics[width=0.35\textwidth]{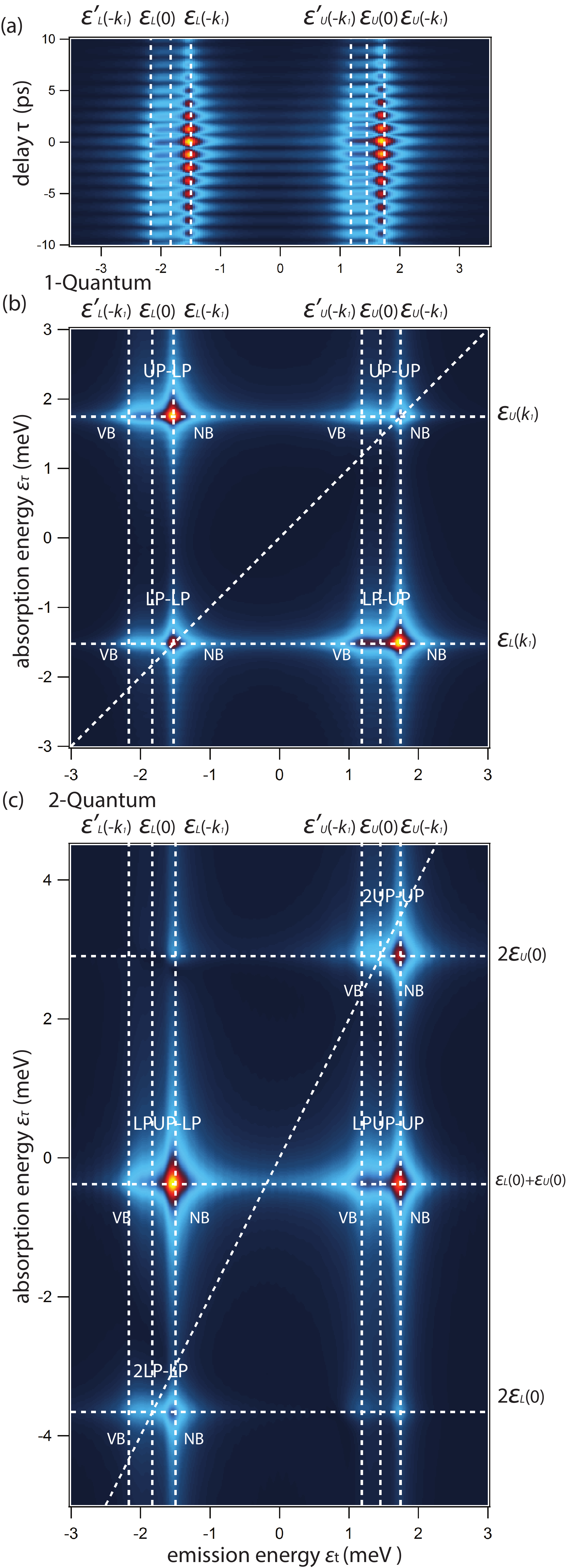}
\caption{Simulation based on exciton-photon (local mode) basis Gross-Pitaevskii-equations. Amplitude of delaymap $|S(\tau,\epsilon_t)|$ (a). Amplitude of 2D FWM spectrum $|S(\epsilon_\tau,\epsilon_t)|$ for 1-quantum (b) and 2-quantum region (c).}
\label{fig:localmode}
\end{center}
\end{figure}    
The first term is a linear term of the unperturbed energy $E_{n,m}$. The next three terms (a'1)-(a'3) are the first-order perturbation terms resulting from the LP-LP (UP-UP) self and LP-UP cross-interaction terms. The remaining parts are the second-order perturbation terms. In the main text, we neglect the second-order terms (b'1)-(c'2). Let us evaluate the condition where this approximation holds assuming that the lower and upper-polariton densities ($m=n$) are equal and the cavity detuning is zero ($|X|=|C|$). The calculated energy indicates that the first-order perturbation term is proportional to $g_0n^2$, while the second-order one is to $g_0n^4/\Omega$. Thus, the approximation which takes into account the first three terms is satisfied under the condition: $g_0n^2/\Omega<1$. Hence, a large Rabi splitting and low polariton density is required for our approximation to hold. Actually, if the number of polaritons becomes large, this perturbation breaks down and the contribution from the terms (b1)-(c2) becomes comparable to (a1)-(a2). For sake of rigor, we note that our treatment only holds when the polariton-polariton scattering terms away from q=0 can be ignored.

In addition to the above perturbative discussion, we present 2DFT spectra based on exciton-photon Hamiltonian Eq. \ref{ecHamiltonian} using numerical simulations. Similarly to the polariton Gross-Pitaevskii equations, with a mean-field approximation, we can derive conventional exciton-photon Gross-Pitaevskii equations \cite{Kohnle2012}:
\begin{eqnarray}
i\hbar \dot{\psi}_{x}&=&(\epsilon_{x}-\frac{\hbar^2}{2m_x}\nabla^2+g_{0}|\psi_{x}|^2-i\frac{\gamma_x}{2})\psi_{x}\\
i\hbar \dot{\psi}_{c}&=&(\epsilon_{c}-\frac{\hbar^2}{2m_c}\nabla^2-i\frac{\gamma_{c}}{2})\psi_{c}-f_{\rm ext}.
\end{eqnarray}
The simulation of the cavity detuning $\delta$=-0.38 meV is presented in Fig. \ref{fig:localmode}. The parameters used in this simulation is same as that of polariton-based calculation performed in the main text (Fig. \ref{fig:sim_PRB}). In this exciton-photon  basis calculation, both nonparabolicity of polariton's energy-momentum dispersion and all polariton-polariton interaction terms Eq (a1)-(c2) are automatically taken into account. The comparison between exciton-photon (Fig. \ref{fig:localmode}) and polariton basis calculations (Fig. \ref{fig:sim_PRB}) show that both frameworks give qualitatively the same results for low polariton densities. In the high density regime, the results of the two simulations change (not shown), where the lower-polariton basis calculation breaks down, while the exciton-photon basis Gross-Pitaevskii still works.

%\bibliography{bib2DFWM}

\begin{thebibliography}{32}
\bibitem [{foo()}]{footnote}%
  \BibitemOpen
  \href@noop {} {\bibinfo  {journal} {{Strictly speaking, the origin of the
  appearance of peaks in 2DFT spectrum is not restricted to an energy
  blue-shift. Any kind of nonlinearity such as density dependent decay
  (excitation induced dephasing) is also possible to produce a FWM signal. In
  this article, for simplicity of the discussion, we restrict nonlinearities to
  the mean-field blue shift due to polariton-polariton interactions}}\
  }\BibitemShut {NoStop}%
\bibitem [{a()}]{a}%
  \BibitemOpen
  \href@noop {} {\bibinfo  {journal} {{Photon-assisted exchange scattering can
  be included by adding a term $\hat{H}_{\rm PAE}=g_{\rm PAE}(\hat{\bm
  \psi}_{c}^{\dagger}\hat{\bm \psi}_{x}^{\dagger}\hat{\bm \psi}_{x}\hat{\bm
  \psi}_{x}+\hat{\bm \psi}_{x}^{\dagger}\hat{\bm \psi}_{x}^{\dagger}\hat{\bm
  \psi}_{x}\hat{\bm \psi}_{c})$}. In several articles, this term is referred to
  as ``phase-space filling"}\ }\BibitemShut {NoStop}%
\end{thebibliography}

\end{document}